\documentclass[prd,twocolumn,nofootinbib,showpacs]{revtex4-1}
\usepackage{feynmp,slashed,color,mathrsfs,nicefrac,exscale}
\usepackage[colorlinks=true,linkcolor=blue,citecolor=blue,urlcolor=blue]{hyperref}

\usepackage{epsfig}
\usepackage[colorlinks=true,linkcolor=blue,citecolor=blue,urlcolor=blue]{hyperref}

\usepackage{amsmath}
\usepackage{amssymb}
\usepackage{marginnote}
\usepackage{braket}
\usepackage{xcolor}
\usepackage{mathtools}

\def\beq{\begin{equation}}
\def\eeq{\end{equation}}
\def\beqa{\begin{eqnarray}}
\def\eeqa{\end{eqnarray}}
\def\bea{\begin{eqnarray}}
\def\eea{\end{eqnarray}}
\def\nn{\nonumber}

\def\ss{\scriptscriptstyle}

\begin{document}

\title{Strong couplings and form factors of charmed mesons in holographic QCD}

\author{Alfonso Ballon-Bayona} \email{aballonb@ift.unesp.br}

\author{Gast\~ao Krein} \email{gkrein@ift.unesp.br}

\author{Carlisson Miller} \email{miller@ift.unesp.br}
\affiliation{Instituto de F\'{\i}sica Te\'orica, Universidade Estadual
  Paulista,  Rua Dr. Bento Teobaldo Ferraz, 271 - Bloco II,
  01140-070 S\~ao Paulo, SP, Brazil}

\begin{abstract}
We extend the two-flavor hard-wall holographic model of Erlich, Katz, Son and Stephanov [Phys.\ Rev.\ Lett.\
{\bf 95}, 261602 (2005)] to four flavors to incorporate strange and charm quarks. The model incorporates chiral
and flavor symmetry breaking and provides a reasonable description of masses and weak decay constants
of a variety of scalar, pseudoscalar, vector and axial-vector strange and charmed mesons. In particular, we examine flavor
symmetry breaking in the strong couplings of the $\rho$ meson to the charmed $D$ and $D^{*}$ mesons.
We also compute electromagnetic form factors of the $\pi$,  $\rho$, $K$, $K^*$, $D$ and $D^*$ mesons. We compare our results for the $D$ and $D^*$ mesons
with lattice QCD data and other nonperturbative approaches.
\end{abstract}

\maketitle

%%%%%%%%%%%%%%%%%%%%%%%%%%%%%%%%%%%%%%%%%%%%%%%%%%%%%%%%%%%%%%%%%%%%%%%%%%%%%%%%%%%%%%%%%%%
\section{Introduction}

There is considerable current theoretical and experimental interest in the study of the interactions
of charmed hadrons with light hadrons and atomic nuclei~\cite{{Krein:2016fqh},{Briceno:2015rlt},
{Hosaka:2016ypm}}. There is special interest in the properties of $D$ mesons in nuclear matter~\cite{Tolos:2013gta},
mainly in connection with $D-$mesic nuclei~\cite{{Tsushima:1998ru},{Yasui:2009bz},{GarciaRecio:2010vt},
{GarciaRecio:2011xt}}, $J/\Psi$ and $\eta_c$ binding to nuclei~\cite{{Brodsky:1989jd},{Ko:2000jx},{Krein:2010vp},
{Tsushima:2011kh}}, and $ND$ molecules~\cite{Carames:2016qhr}. $D-$mesons are also of interest in the context of the
so-called X,Y,Z exotic hadrons, which have galvanized the field of hadron spectroscopy since the discovery in 2003
of the charmed hadron $X(3872)$ by the Belle collaboration~\cite{Choi:2003ue}. They are {\em exotic} because they
do not fit the conventional quark-model pattern of either quark-antiquark mesons or three-quark baryons. Most of
the X,Y,Z hadrons have masses close to open-flavor thresholds and decay into hadrons containing charm (or bottom)
quarks. Presently there is no clear theoretical understanding of the new hadrons, despite of the huge literature
that has accumulated over the last decade. In the coming years, existing and forthcoming experiments will produce
numerous new, and very likely surprising results{\textemdash}Ref.~\cite{Lebed:2016hpi} is
a very recent review on exotic hadrons, with an extensive list of references on theory and experiment.
The $\overline{\rm P}$ANDA collaboration~\cite{PANDA}, in particular, at the forthcoming FAIR facility has an
extensive program~\cite{{Wiedner:2011},{Prencipe:2016}} aiming at the investigation of charmed hadrons and
their interactions with ordinary matter.

A major difficulty in the theoretical treatment of in-medium interactions of charmed hadrons is the lack
of experimental information on the interactions in free space. For example, almost all knowledge on the $DN$
interaction comes from calculations based on effective Lagrangians that are extensions of light-flavor
chiral Lagrangians using $SU(4)$ flavor symmetry~\cite{{Mizutani:2006vq},{Lin:1999ve},{Hofmann:2005sw},
{Haidenbauer:2007jq},{Haidenbauer:2008ff},{Haidenbauer:2010ch},{Fontoura:2012mz}} and heavy quark
symmetry~\cite{{Yasui:2009bz},{GarciaRecio:2008dp}}. The Lagrangians involve coupling constants, like
$g_{\rho{DD}}$, $g_{\omega{DD}}$, $g_{\rho{D^\ast D}}$ and $g_{\rho{D^\ast D^\ast}}$, whose values are taken
from $SU(4)$ flavor and heavy-quark symmetry relations. For instance, $SU(4)$ symmetry relates the couplings of
the $\rho$ to the pseudoscalar mesons $\pi$, $K$ and $D$, namely $g_{\rho{DD}} = g_{KK\rho} = g_{\rho{\pi\pi}}/2$.
If in addition to $SU(4)$ flavor symmetry, heavy-quark spin symmetry is invoked, one has $g_{\rho DD} =
g_{\rho D^\ast D} = g_{\rho D^\ast D^\ast} = g_{\pi D^\ast D}$ to leading order in the charm quark
mass~\cite{Casalbuoni:1996pg, Manohar:2000dt}. The coupling $g_{\rho{\pi\pi}}$ is constrained by experimental
data; the studies of the $DN$ interaction in
Refs~\cite{Haidenbauer:2007jq,Haidenbauer:2008ff,Haidenbauer:2010ch} utilized such a $SU(4)$ relation, taking
$g_{\rho\pi\pi} = 6.0$, which is the value used in a large body of work conducted within the J\"ulich
model~\cite{{Haidenbauer:1991kt},{Hoffmann:1995ie}} for light-flavor hadrons. This value of $g_{\rho\pi\pi}$
implies through $SU(4)$ symmetry $g_{\rho DD} = 3$, which is not very much different from predictions based on
the vector meson dominance (VMD) model: $g_{DD\rho} = 2.52-2.8$~\cite{Mat98,Lin00a}. Moreover, to maintain
unitarity in calculations of scattering phase shifts and cross sections, lowest-order Born diagrams need to be
iterated with the use of a scattering equation, like the Lippmann-Schwinger equation, and phenomenological
form factors are required to control ultraviolet divergences. Form factors involve cutoff parameters that
also are subject to flavor dependence. Again, due to the lack of experimental information, they
are also poorly constrained.

Flavor symmetry is strongly broken at the level of the QCD Lagrangian due to the widely different values of the
quark masses; while in the light quark sector one has good $SU(2)$ symmetry, $m_u \simeq m_d$, thereby e.g.
$g_{\rho{DD}} = g_{\omega{DD}}$ (up to a phase), in the heavy-flavor sector $SU(3)$ and $SU(4)$ symmetries are
badly broken: $m_c \gg m_s \gg m_u$. Given the importance of effective Lagrangians in the study of a
great variety of phenomena involving $D-$mesons, in the present examine their properties in a holographic model
of QCD. We extend the holographic QCD model of  Refs.~\cite{Erlich:2005qh,Da Rold:2005zs} to the case of $N_f=4$
and investigate the implications of the widely different values of the quark masses on the effective three-meson
couplings $g_{\rho D D}$ and $g_{\rho D^* D^*}$ and the electromagnetic form factors of the $D$ and $D^*$ mesons.
The parameters of the model are the quark masses and condensates as well as the mass gap scale.  Using experimental
data for a selected set of meson masses to fix the model parameters, allows us to predict not only
the strong couplings and electromagnetic form factors mentioned above but also many other observables not
studied before with a holographic model.

The works in Refs.~\cite{Erlich:2005qh,Da Rold:2005zs} pioneered in the modeling of low energy QCD by incorporating
features of dynamical chiral symmetry breaking in holographic QCD. They correctly identify the five dimensional
gauge fields dual to the left and right currents
associated with chiral symmetry as well as the five dimensional scalar field dual to the chiral condensate.
The extension proposed in Ref.~\cite{Abidin:2009aj} incorporated the strange quark and was able to identify
the appearance of scalar modes associated with flavor symmetry breaking. In the present work, by extending the
model of Refs.~\cite{Erlich:2005qh,Da Rold:2005zs} to the case $N_f=4$, we are able to investigate the
consequences of the dramatically different values of the quark masses on the phenomenology of charmed mesons.
Moreover, by combining the formalism of Kaluza-Klein expansions and the AdS/CFT dictionary, we are able to directly
extract the leptonic decay constants of mesons and find an expansion for the flavor currents that relate flavor
symmetry breaking to the appearance of scalar modes. That relation bears a strong analogy with the generalized
PCAC (partially conserved axial current relation)~\cite{{Holl:2004fr},{Dominguez:1976ut}} that relates dynamical
chiral symmetry breaking to the appearance of the pion and its resonances~\cite{Ballon-Bayona:2014oma}.

In the model of  Refs.~\cite{Erlich:2005qh,Da Rold:2005zs}, dynamical chiral symmetry breaking becomes manifest
when considering fluctuations of the five dimensional gauge fields associated with the axial and vectorial sector.
While the kinetic terms of the axial sector acquire a mass, signalizing chiral symmetry breaking, the vector
sector remains massless. In our framework, it turns out that the vector sector also acquires a mass signalizing
the breaking of flavor symmetry. The Kaluza-Klein decomposition of these fields  allows us to obtain effective
kinetic Lagrangians for the mesons from the five dimensional kinetic terms, with masses and decay constants
obtained in terms of the wave functions representing the Kaluza-Klein modes. Moreover, expanding the five
dimensional action to cubic order in the fluctuations and performing again a Kaluza-Klein decomposition allows
us to obtain effective Lagrangians describing the three-meson interactions, with strong couplings given in terms
of integrals involving the wave functions of the corresponding Kaluza-Klein modes.

It turns out that the symmetry breaking pattern in the strong couplings differs somewhat from previous studies
in the literature. Calculations employing QCD sum rules found $SU(4)$ symmetry breaking in three-hadron couplings
that vary within the range of 7\% to 70\%\cite{Bracco:2011pg}. In Ref.~\cite{ElBennich:2011py}, using a model constrained by the Dyson-Schwinger equations of QCD, it was found that the relation $g_{\rho DD} =
g_{\rho \pi\pi}/2$ is strongly violated at the level of 300\% or more. In a recent follow up of that study
within the same framework, Ref.~\cite{El-Bennich:2016bno} finds that couplings between $D$-, $D^\ast$-mesons
and $\pi$-, $\rho$-mesons can differ by almost an order-of-magnitude, and that the corresponding form factors
also exhibit different momentum dependences. Our results calculations are more in line with calculations
using the  $^3{\rm P}_0$ quark-pair creation model in the nonrelativistic quark model~\cite{{Krein:2012lra},{Fontoura:2017ujf}}.

The organization of this paper is as follows. In Sec.~\ref{Sec:CSBFSB} we describe how chiral and flavor
symmetry breaking is realized in our model. Then in Sec.~\ref{Sec:Currents} we describe the five dimensional
field equations and the AdS/CFT dictionary for the flavor and axial currents. In Sec.~\ref{Sec:kinetic} we
describe the formalism of Kaluza-Klein expansions and obtain effective kinetic Lagrangians for the mesons.
In Sec.~\ref{Sec:Decayconstants} we use the prescription of our previous studies in Ref.~~\cite{Ballon-Bayona:2014oma}
for the leptonic decay constants and obtain relations describing flavor symmetry breaking and chiral symmetry
breaking in terms of scalar and pseudoscalar modes respectively. In Sec.~\ref{Sec:Couplings} we obtain
effective Lagrangians describing three-meson interactions with the holographic prescription for the strong
couplings. Finally, in Sec.~\ref{Sec:Numerics} we fit the model parameters and present our numerical results
for many observables, including the strong couplings $g_{\rho DD}$ and $g_{\rho D^* D^*}$ as well as
the electromagnetic form factors of the $D$ and $D^*$ mesons. We compare the latter against lattice QCD
data obtained in Ref.~\cite{Can:2012tx}. Section~\ref{Sec:concl} presents our conclusions.

%%%%%%%%%%%%%%%%%%%%%%%%%%%%%%%%%%%%%%%%%%%%%%%%%%%%%%%%%%%%%%%%%%%%%%%%%%%%%%%%%
%
\section{Chiral symmetry and flavor symmetry in holographic QCD}
\label{Sec:CSBFSB}

Chiral symmetry $SU(N_f)_L \times SU(N_f)_R$ for $N_f$ flavors holds in the massless limit of QCD
and is described in terms of the left and right currents
\beqa
J_{L/R}^{\mu , a} = \bar q_{L/R} \gamma^\mu T^a q_{L/R} \, ,
\eeqa
where $T^a$, $a=1, \dots N^2_f-1$ are the generators of the $SU(N_f)$ group,
and $q_{L/R} = 1/2\,(1\pm\gamma_5) q$, with $q$ being the quark Dirac field. The $SU(4)$ generators
$T^a$ are normalized by the trace condition ${\rm Tr} ( T^a T^b ) = 1/2 \, \delta^{ab}$, satisfying the
Lie algebra $\left [ T^a , T^b \right ] = i f^{abc} T^c$. The generators $T^a$ are related to the
Gell-Mann matrices $\lambda^a$ by $T^a = 1/2 \,\lambda^a$. Chiral symmetry is broken by the presence
of the operator $\bar q q = \bar q_R q_L + \bar q_L q_R$. This breaking can be
explicit, when it appears in the QCD Lagragian associated with the nonzero quark masses, or dynamically, when
it acquires a vacuum expectation value, giving rise to a condensate $\langle \bar q q\rangle$ in limit of
zero quark masses.

In the case $N_f=2$, dynamical chiral symmetry breaking goes as $SU(2)_L \times SU(2)_R \to SU(2)_V$,
where $SU(2)_V$ is an exact vector symmetry and the broken symmetry occurs in the axial sector. This is described
in terms of the vector and axial currents $J^{\mu,a}_{V/A}= J^{\mu,a}_R \pm J^{\mu , a}_L$. The symmetry
associated with the vector sector $J^{\mu,a}_V$ is known as isospin symmetry. When $N_f >2$, both chiral and
flavor symmetries are broken by the quark masses. We will describe how chiral and flavor symmetry breaking are
implemented in a holographic model for $N_f=4$.

In the pioneering work of Refs.~\cite{Erlich:2005qh} and \cite{Da Rold:2005zs}, a simple holographic realization
of chiral symmetry breaking (CSB) was proposed. They considered the simplest background in holographic QCD,
known as the hard wall model~\cite{Polchinski:2001tt}, consisting in a slice of anti-de-Sitter spacetime:
\beqa
ds^2 = \frac{1}{z^2} \left ( \eta_{\mu \nu} dx^{\mu} dx^{\nu} - dz^2   \right )  \, , \label{AdSMetric}
\eeqa
with $0 < z  \le z_0$. The parameter $z_0$ determines an infrared (IR) scale at which conformal symmetry
is broken. The action proposed in Ref.~\cite{Erlich:2005qh} includes $N_f$ gauge fields $L_m$ and $R_m$,
corresponding to the left and right flavor currents $J_{L/R}^{\mu , a}$, and a bifundamental field $X$ dual
to the operator $\bar q_R q_L$.  The action can be written as
\beqa
S &=& \int d^5 x \sqrt{|g|} {\rm Tr} \Big \{ (D^m X)^{\dag} (D_m X)  + 3 |X|^2 \cr
&-& \frac{1}{4 g_5^2} \left ( L^{mn} L_{mn} + R^{mn} R_{mn}  \right )
\Big \} \, ,  \label{Erlichaction}
\eeqa
where $D_m X = \partial_m X - i L_m X + i X R_m$ is the covariant derivative of the bifundamental field $X$,
and
\beqa
L_{mn} &=& \partial_m L_n - \partial_n L_m - i \left [ L_m , L_n \right ], \nn \\
R_{mn} &=& \partial_m R_n - \partial_n R_m - i \left [ R_m , R_n \right ],
\eeqa
are non-Abelian field strengths. The 5-d squared mass of the field $X$ is fixed to  $m^2 = -3$, to match
with the conformal dimension $\Delta=3$ of the dual operator $\bar q_R q_L$. The model of Ref.~\cite{Erlich:2005qh}
focused on $N_f=2$ and worked in the limit of exact flavor symmetry. In Ref.~\cite{Abidin:2009aj},
Abidin and Carlson extended the model to $N_f=3$, to incorporate the strange-quark sector. In the present
paper we further extend that model to $N_f=4$, with the aim of making predictions for charmed mesons. In our
approach we use a Kaluza-Klein expansion for the 5-d fields in order to find a 4-d effective action for the
mesons. This approach allows us to find directly the meson weak decay constants, couplings and expansions
for the vector and axial currents. We find in particular a relation for the vector
current describing flavor symmetry breaking (FSB).

We start with the classical background that describes chiral symmetry breaking:
\beqa
L^0_m = & R^0_m = 0 \, , \quad 2 X_0 = \zeta M z + \frac{\Sigma}{\zeta} z^3 ,
\label{classicalfields}
\eeqa
where $M$ is the quark-mass matrix, $M={\rm diag}(m_u , m_u , m_s , m_c)$, and $\Sigma$ is the matrix of the
quark condensates, $\Sigma={\rm diag}(\sigma_u ,  \sigma_u , \sigma_s , \sigma_c)$. The parameter $\zeta=\sqrt{N_c}/2\pi$  is introduced to have consistency with the counting rules of large-$N_c$  QCD{\textemdash}for details, see Ref.~\cite{Cherman:2008eh}. Note that we are assuming $SU(2)$ isospin symmetry in the light-quark sector, i.e. $m_d = m_u$ and $\sigma_d = \sigma_u$, which is a very good approximate symmetry in QCD.

For the strange and charm quarks we will fit their masses $m_s$ and $m_c$ to the physical masses for the mesons. Note, however, that the model should not be valid for arbitrarily large quark masses. The reason is that, from the
string theory perspective, the action in equation (\ref{Erlichaction}) is expected to arise from a small perturbation
of $N_f$ coincident space-filling flavor branes. Specifically, the mass term $M$ appearing in
Eq.~(\ref{classicalfields}) acts as a small source for the operator $ \bar q_R q_L$, responsible for the
breaking of the chiral  and flavor symmetries. A holographic description of quarks with very large masses requires
the inclusion of long open strings and two sets of flavor branes distinguishing the heavy quarks from the
light quarks (see e.g. \cite{Erdmenger:2006bg}). In that framework, the string length is proportional to the
quark mass and each set of flavor branes will carry a set of fields describing the dynamics of light and
heavy mesons respectively. In this work we will show that the model described by the action in
equation (\ref{Erlichaction}) is still a very good approximation for the dynamics of light and
heavy-light charmed mesons, the reason being that the internal structure in both cases is governed by
essentially the same nonperturbative physics, that occurs at the scale $\Lambda_{\rm QCD}$~\cite{Manohar:2000dt}.
In heavy-heavy mesons, on the other hand, the internal dynamics is governed by short-distance
physics. For recent holographic studies of mesons involving heavy quarks see
Refs.~\cite{Hashimoto:2014jua,Braga:2015lck,Liu:2016iqo,Liu:2016urz}.

To investigate the consequences of chiral and flavor symmetry breaking it is convenient to rewrite the fluctuations
of left and right gauge fields in terms of vector and axial fields , i.e. $L_m = V_m + A_m$ and
$R_m = V_m - A_m$.  The bifundamental field X can be decomposed as
\beq
X = e^{i\pi} \, X_0 \, e^{i\pi},
\eeq
where $X_0$ is the classical part and $\pi$ contains the fluctuations. The fields $V_m$, $A_m$ and $\pi$
can be expanded as $V_m^a T^a$, $A_m^a T^a$ and $\pi^a T^a$ respectively.

It is important to remark that organizing the heavy-light D mesons together with the light pions and kaons in a 15-plet $\pi^a T^a$ of fluctuation  
fields does not imply, automatically, that the heavy-light D mesons are being approximated by Nambu-Goldstone bosons. The reason is that the explicit breaking of chiral symmetry, driven by the 
heavy charm quark, is large and by no means its effects are neglected in the model. In the same way, 
the fact the $D$ mesons appear in the same multiplet of the SU(4) flavor group does not mean that flavor symmetry is exact; it is explicitly broken by the widely different values of the quark masses. The 
main advantage of using such a SU(4) representation with explicit symmetry-breaking terms is that it 
allows us to make contact with the four dimensional effective field theories describing the interactions
of light and heavy-light mesons commonly used in phenomenological applications. This not only extends 
the work of  \cite{Lin:1999ve,Lin00a} but also leads to quantitative predictions for the strong couplings that can 
be tested against experiment or lattice QCD data, which is our main objective in the present paper. 
An alternative approach to describe the heavy-light mesons is to make contact with a particularly interesting class of four dimensional models that treat
the light mesons as in the present paper, and treat heavy mesons by invoking heavy-quark symmetry. The Lagrangian
in the heavy sector is written as an expansion in inverse powers of the heavy quark mass; Refs. \cite{Yasui:2009bz,GarciaRecio:2008dp} are examples of such models. In holography the heavy quarks are realized in terms of long open strings, as described above in this section. For recent progress in the heavy quark approach to heavy-light mesons within holographic QCD see \cite{Liu:2016iqo,Liu:2016urz}.

Expanding the action in Eq.~(\ref{Erlichaction}) up to cubic order in the fields $V_m^a$, $A_m^a$ and $\pi^a$,
we find
\beq
S = S^{(2)} + S^{(3)} + \dots \, ,
\eeq
where
\begin{widetext}
\beqa
S^{(2)} &=&  \int d^5 x \sqrt{|g|} \Big \{ - \frac{1}{4 g_5^2} v^{mn}_a v_{mn}^a + \frac{1}{2} \left(M^a_V\right)^2
V^m_a V_m^a - \frac{1}{4 g_5^2} a^{mn}_a a_{mn}^a
+ \frac{ M_A^{a b}}{2} (\partial^m \pi^a - A^{m,a}) (\partial_m \pi_b - A_{m,b}) \Big \}
\label{S2}
\eeqa
\beqa
S^{(3)} &=&  \int d^5 x \sqrt{|g|} \Big \{ - \frac{1}{2g_5^2} f^{abc}  v^{mn}_a \left ( V_m^b V_n^c + A_m^b A_n^c \right )
- \frac{1}{g_5^2} f^{abc} a^{mn}_a V_m^b A_n^c
 - \frac{(M_V^b)^2}{2} f^{abc} \,  (\partial_m \pi^a - 2 A_m^a) V^{m,b} \pi^c  \cr
 &+& M_A^{ae} f^{ebc} \,  (\partial_m \pi^a - A_m^a) V^{m,b} \pi^c \Big \} \, ,  \label{S3}
\eeqa
\end{widetext}
and we have defined the Abelian field strengths $v_{mn}^a = \partial_m V_n^a - \partial_n V_m^a$
and $a_{mn}^a = \partial_m A_n^a - \partial_n A_m^a$. In the kinetic term $S^{(2)}$, the vector and axial
symmetry breaking is dictated by the mass terms $M_V^a$ and $\tilde M_A^{a b}$,
defined by the traces $2 {\rm Tr} ( [T^a , X_0] [ T^b, X_0] )  = - (M_V^a)^2 \delta^{ab}$ and $2 {\rm Tr} ( \{ T^a , X_0 \} \{ T^b , X_0 \} ) =  M_A^{ab}$. Note, however, that the axial sector in $S^{(2)}$ is invariant
under the gauge transformation
\beqa
A^{a}_{m} \rightarrow A^{a}_{m}-\partial_{m}\lambda_A^{a} \quad , \quad
 \pi^{a} \rightarrow \pi^{a} - \lambda_A^{a} \, . \label{Axialsym}
\eeqa

Using Eq.~(\ref{classicalfields}) we find the following nonzero values for $M_V$:
\beqa
&&
(M_V^a)^2 =   \frac14 (v_s - v_u)^2  \quad {\rm for} \, \, a= (4,5,6,7) \, , \cr
&&
(M_V^a)^2 =   \frac14 (v_c - v_u)^2  \quad {\rm for} \, \, a= (9,10,11,12) \, , \cr
&&(M_V^a)^2 =   \frac14 (v_c - v_s)^2  \quad {\rm for} \, \, a= (13,14) \, , \label{MV}
\eeqa
and the nonzero values for $M_A$:
\beqa
&&M_A^{a,a} = v_u^2  \qquad \qquad \quad {\rm for} \, \, a= (1,2,3) \, , \cr
&& M_A^{a,a} = \frac14 (v_s + v_u)^2\quad {\rm for} \, \, a= (4,5,6,7) \, , \cr
&& M_A^{a,a} =  \frac14 (v_c + v_u)^2 \quad {\rm for} \, \, a= (9,10,11,12) \, , \cr
&& M_A^{a,a} = \frac14 (v_c + v_s)^2 \quad {\rm for} \, \, a= (13,14) \, , \cr
&&M_A^{8,8} = \frac13 (v_u^2 + 2 v_s^2 ) \, , \cr
&& M_A^{15,15} = \frac{1}{12} (2 v_u^2 + v_s^2 + 9 v_c^2 )  \,,  \cr
&& M_A^{8,15} = M_A^{15,8} = \frac{1}{3 \sqrt{2}} (  v_u^2  -  v_s^2) \, .  \label{MA}
\eeqa
In Eqs.~(\ref{MV}) and (\ref{MA}) we have defined
\beqa
v_q (z) = \zeta m_q z + \frac{1}{\zeta} \sigma_q z^3 \, \,  , \, \,  q=(u,s,c) \,.
\eeqa

In the interesting case where all the masses and condensates are equal we have that $(M_V^a)^2=0$
and the $SU(4)$ flavor symmetry is preserved. In this paper we consider quark masses and condensates that lead to a realistic
spectrum for the mesons so that we could explore the consequences of $SU(4)$  flavor symmetry breaking. The kinetic term in Eq.~(\ref{S2}) allows us to  extract the meson spectrum and decay
constants whereas the action in Eq.~(\ref{S3}) leads to nontrivial predictions for three-meson couplings,
including the heavy-light charmed mesons $D$ and $D^{*}$.

Note that the mass term for the vectorial sector $(M_V^a)^2$ is zero not only for $a=(1,2,3)$, corresponding to the light  $SU(2)$ sector but also for $a=(8,15)$, which implies that flavor symmetry has not been broken in the sector describing the dynamics of the $\omega'$ and $\psi$ mesons. This is one clear example of heavy-heavy mesons (mesons composed by a heavy quark-antiquark pair), where we actually expect some corrections to appear in (\ref{Erlichaction}) describing flavor symmetry breaking. Those terms would arise from the dynamics of long open strings dual to heavy quarks, as explained above in this section.

\section{Field equations and dual currents}

\label{Sec:Currents}

Writing the kinetic action in Eq.~(\ref{S2}) as $S^{(2)} = \int d^5 x \, {\cal L}^{(2)}$, its variation
takes the form $\delta S^{(2)} =
\delta S^{(2)}_{\rm Bulk} + \delta S^{(2)}_{\rm Bdy}$ where
\beqa
\delta S^{(2)}_{\rm Bulk}  &=& \int d^5 x \, \Biggl [ \left ( \frac{\partial {\cal L}^{(2)}}
{\partial V_\ell^a} - \partial_m P_{V , a} ^{m \ell}\right ) \, \delta V_{\ell}^a
 \cr
 &+& \left ( \frac{\partial {\cal L}^{(2)}} {\partial A_\ell^a}
- \partial_m P_{A , a} ^{m \ell}\right ) \, \delta A_{\ell}^a \cr
&+& \left ( \frac{\partial {\cal L}^{(2)} } {\partial \pi^a}
- \partial_m P_{\pi , a}^m \right ) \, \delta \pi^a \Biggr] \, ,
\label{deltaS2Bulk}
\eeqa
\beqa
\delta S^{(2)}_{\rm Bdy} = \int d^5 x \, \partial_m \Big ( P_{V , a}^{m \ell}
\delta V_{\ell}^a + P_{A , a}^{m \ell} \, \delta A_{\ell}^a + P_{\pi , a}^m \, \delta \pi^a \Big ) \, , \cr
&&  \label{deltaS2Bdy}
\eeqa
and
\beqa
P_{V , a} ^{m \ell} &:=&
\frac{\partial {\cal L}^{(2)}}{ \partial (\partial_m V_\ell^a) } = - \frac{1}{g_5^2} \sqrt{|g|} \,  v^{m \ell}_a  \, , \cr
P_{A , a} ^{m \ell} &:=&
\frac{\partial {\cal L}^{(2)}}{ \partial (\partial_m A_\ell^a) } = - \frac{1}{g_5^2} \sqrt{|g|} \,  a^{m \ell}_a \, , \cr
P_{\pi , a}^m &:=&
\frac{\partial {\cal L}^{(2)}}{\partial ( \partial_m \pi^a ) } = M_A^{ab} \sqrt{|g|} \, (\partial^m \pi_b - A^m_b ) \, ,
\eeqa
are conjugate momenta to the 5-d fields $V_m^a$, $A_m^a$ and $\pi^a$ respectively.
%and
%\beqa
%\frac{\partial {\cal L}^{(2)}}{\partial V_\ell^a}  &=&
%(M_V^a)^2 \sqrt{|g|} V^{\ell}_a \, , \cr
% \frac{\partial {\cal L}^{(2)}}{\partial A_\ell^a}  &=& - %M_A^{ab} \sqrt{|g|}
%\left ( \partial^\ell \pi^b - A^{\ell ,b}\right )  \, , \cr
%\frac{\partial {\cal L}^{(2)}}{\partial \pi^a} &=& 0 \, ,
%\eeqa
The bulk term in Eq.~(\ref{deltaS2Bulk}) leads to the field equations
\beqa
&& \partial_m \left [ \sqrt{|g|} v^{mn}_a \right ] + g_5^2 (M_V^a)^2 \sqrt{|g|}\,  V^n_a = 0 \, , \cr
&& \partial_m \left [ \sqrt{|g|} a^{mn}_a \right ] - g_5^2 M_A^{ab} \sqrt{|g|} \, (\partial^n \pi_b - A^n_b ) = 0 \, , \cr
&&
\partial_m \left [ M_A^{ab} \sqrt{|g|} \, (\partial^m \pi_b - A^m_b ) \right ] = 0 \,. \label{FieldEqs}
\eeqa

%The surface term (\ref{deltaS2Bdy}) has a nonvanishing %contribution
%
%\beqa
%\delta S^{(2)}_{\rm Bdy} &=& \int d^4 x  \Bigl( P_{V , a}^{z %\mu} \, \delta V_{\mu}^a
%+ P_{A , a}^{z \mu} \, \delta A_{\mu}^a  \cr
% &+&   P_{\pi , a}^z \, \delta \pi^a \Bigr)^{z=z_0}_{z=%\epsilon}.
%\eeqa
Imposing the boundary conditions  $\partial_z V_\mu^a |_{z=z_0} = \partial_z A_\mu^a |_{z=z_0} = 0 $ and $\partial_z \pi^a |_{z=z_0} = A_z |_{z=z_0} = V_z |_{z=z_0} =  0 $ the boundary term (\ref{deltaS2Bdy}) reduces to
%
%\beqa
%\partial_z V_\mu^a |_{z=z_0} &=& \partial_z A_\mu^a |_{z=z_0} %= 0 \, , \cr
%\partial_z \pi^a |_{z=z_0} &=& A_z |_{z=z_0} = V_z |_{z=z_0} =  %0  \, .
%\eeqa
\beqa
\delta S^{(2)}_{\rm Bdy} &=& - \int d^4 x  \,  \Bigl [ \langle J^{\hat \mu}_{V,a} \rangle
\, \left( \delta V_{\hat \mu}^a \right)_{z = \epsilon}
+  \langle J^{\hat \mu}_{A,a} \rangle  \, \left(\delta A_{\hat\mu}^a\right)_{z = \epsilon} \cr
&+&   \langle J_{\pi,a} \rangle \, \left(\delta \pi^a \right)_{z = \epsilon} \Bigr ] ,
\eeqa
where we find the dual 4-d currents
\beqa
\langle J^{\hat\mu}_{V,a} (x) \rangle &=&  P_{V , a}^{z \mu} \vert_{z = \epsilon}
= - \frac{1}{g_5^2} \left ( \sqrt{|g|} v^{z \mu}_a \right )_{z = \epsilon} ,
\label{JV} \\
\langle J^{\hat\mu}_{A,a} (x) \rangle &=&  P_{A , a}^{z \mu} \vert_{z = \epsilon}
= - \frac{1}{g_5^2} \left ( \sqrt{|g|} a^{z \mu}_a \right )_{z = \epsilon}  ,
\label{JA} \\
\langle J_{\pi,a} (x) \rangle &=&
 P_{\pi , a}^z \vert_{z = \epsilon}
=   \left [ \sqrt{|g|} M_A^{ab} \left ( \partial^z \pi_b - A^z_b \right ) \right ]_{z = \epsilon} \cr
&=&  \partial_{\hat \mu} \langle J^{\hat \mu}_{A,a} (x) \rangle .
\label{JPi}
\eeqa
Note that we distinguish the vector Minkowski indices $\hat \mu$ from the AdS indices $\mu$.
The results in Eqs.~(\ref{JV}), (\ref{JA}) and (\ref{JPi}) define the holographic prescription for
expectation values of the 4-d vector, axial and pion current operators.
Note from Eqs.~(\ref{FieldEqs}) and (\ref{JV}) that $ \partial_{\hat \mu} \langle J^{\hat\mu}_{V,a} (x) \rangle
\neq 0$ when $(M_V^a)^2 \neq 0$, i.e. the vector current is not conserved for those cases. Similarly,
from Eqs.~(\ref{FieldEqs}) and (\ref{JA}), one sees that  $ \partial_{\hat \mu} \langle J^{\hat\mu}_{A,a} (x) \rangle
\neq 0 $ for any $a$ (because $M_A^{ab} \neq 0$), i.e. the axial current is never conserved.

\section{The 4-d effective action}

\label{Sec:kinetic}

After decomposing the vector and axial fields into their $(z,\mu)$ components and evaluating the metric
in Eq.~(\ref{AdSMetric}), the kinetic action in Eq.~(\ref{S2}) takes the form
\beqa
S^{(2)} &=& S^{(2)}_{V} + S^{(2)}_{A} \, ,
\eeqa
where
\beqa
S^{(2)}_{V} &=& \int d^4 x \int \frac{d z}{z} \Big \{  - \frac{1}{4 g_5^2} \left [ (v_{\hat \mu \hat \nu}^a)^2 - 2 (v_{z \hat \mu}^a)^2 \right ] \cr
&+& \frac{(M_V^a)^2}{2 z^2} \left [ (V_{\hat \mu}^a)^2 - (V_z^a)^2 \right ] \Big \} \, , \label{SV}
\eeqa
and
\beqa
S^{(2)}_{A} &=& \int d^4 x \int \frac{d z}{z} \Big \{ - \frac{1}{4 g_5^2} \left [ (a_{\hat \mu \hat \nu}^a)^2 - 2 (a_{z \hat \mu}^a)^2 \right ] \cr
&+& \frac{M_A^{ab}}{2z^2} \Big [ (\partial^{\hat \mu} \pi^a - A^{\hat \mu , a} ) (\partial_{\hat \mu} \pi^b - A_{\hat \mu}^{b}) \cr
 &-& (\partial_z \pi^a - A_z^{a} ) (\partial_z \pi^b - A_{z}^b) \Big ]
\Big \}. \label{SA}
\eeqa

The vector and axial sectors admit a decomposition in irreducible representations of the Lorentz group.
For the vector sector we find
\beqa
V_{\hat \mu}^a &=& V_{\hat \mu}^{\perp, a} + \partial_{\hat \mu}  (\tilde \phi^a - \tilde \pi^a ),
\nn \\
V_z^a &=& - \partial_z \tilde \pi^a, \label{Vz}
%A_{\hat \mu}^a &=& A_{\hat \mu}^{\perp, a} + \partial_{\hat %\mu} \phi^a  \, ,  \label{Amu}
\label{Vmu}
\eeqa
where $\partial_{\hat \mu} V_{\hat \mu}^{\perp, a} = 0$. The 5-d field $V_{\hat \mu}^{\perp, a}$ describes
an infinite tower of 4-d massive spin 1 fields, i.e. the vector mesons, whereas the 5-d fields $\tilde \phi^a$
and $\tilde \pi^a$ describe an infinite tower of massive scalar fields, i.e. scalar mesons associated
with flavor symmetry breaking (FSB).

On the other hand, the gauge symmetry in Eq.~(\ref{Axialsym}) allows us to decompose the axial sector
as
\beqa
A_{\hat \mu}^a &\to& A_{\hat \mu}^{\perp , a} \quad , \quad
A_z^a \to - \partial_z \phi^ a \, , \cr
\pi^a &\to& \pi^a - \phi^ a \, , \label{AGauge}
\eeqa
where $\partial_{\hat \mu} A_{\hat \mu}^{\perp, a} = 0$. This time the 5-d field $A_{\hat \mu}^{\perp , a}$
will describe an infinite tower of 4-d massive axial spin 1 fields, i.e. the axial vector mesons. The 5-d fields
$\phi^a$ and $\pi^a$ will describe an infinite tower of 4-d pseudoscalar fields, i.e the pions associated with
chiral symmetry breaking (CSB).

Using Eqs.~(\ref{Vmu}), (\ref{Vz}) and (\ref{AGauge})  the actions in Eqs.~(\ref{SV}) and (\ref{SA}) take
the form
\beqa
S^{(2)}_{V} &=& \int d^4 x \int \frac{dz}{z}
 \Big \{ - \frac{1}{4g_5^2} \Big [ (v_{\hat \mu \nu}^{\perp,a})^2 - 2 (\partial_z V_{\hat \mu}^{\perp,a})^2
 \cr
 &-& 2 (\partial_z \partial_{\hat \mu} \tilde \phi^a )^2 \Big ] + \frac{(M_V^a)^2}{2 z^2}
 \Big [ ( V_{\hat \mu}^{\perp, a} )^2 + (\partial_{\hat \mu} \tilde \pi^a - \partial_{\hat \mu} \tilde \phi^a )^2 \cr
 &-& (\partial_z \tilde \pi^a)^2 \Big ]
 + \partial_{\hat \mu} (\dots) \Big \} , \label{SV2}
\eeqa
\beqa
S^{(2)}_{A} &=& \int d^4 x \int \frac{dz}{z} - \frac{1}{4g_5^2} \Big [ (a_{\hat \mu \nu}^{\perp,a})^2
- 2 (\partial_z A_{\hat \mu}^{\perp,a})^2 \cr
&-& 2 (\partial_z \partial_{\hat \mu} \phi^a )^2 \Big ]
+ \frac{M_A^{ab}}{2 z^2}
\Big [ A_{\perp}^{\hat \mu ,a} A_{\hat \mu}^{\perp, b} \cr &+& ( \partial^{\hat \mu} \pi^a - \partial^{\hat \mu} \phi^a) ( \partial_{\hat \mu} \pi^b - \partial_{\hat \mu} \phi^b) - (\partial_z \pi^a) (\partial_z \pi^b) \Big ]
\cr &+& \partial_{\hat \mu} (\dots) \Big \}, \label{SA2}
\eeqa
where the terms in $(\dots)$ are surface terms that vanish after choosing periodic boundary conditions for the fields.

The actions in Eqs.~(\ref{SV2}) and (\ref{SA2}) are in a suitable form to perform a Kaluza-Klein expansion for the
5-d fields. Before performing this expansion note that the nondiagonal mass term $M_A^{8,15}$ induces a
mixing in the axial sector for meson states with flavor indices  $a=(8,15)$. In this paper we are mainly
interested in the axial sector states with $a=(9,..,14)$ and $a=(1,2,3)$, corresponding to the heavy-light
charmed mesons and the usual light mesons. Then from now on we will consider for the axial sector only those states
where $a \neq (8,15)$. The axial sector states corresponding to $a=(8,15)$ have an interesting physical
interpretation, e.g $\eta - \eta_c$ mixing for the pseudoscalar sector, and deserve a further study that
will be pursued in a future project.

The 5-d fields in the vector sector admit a Kaluza-Klein expansion of the form
\beqa
V_{\hat \mu}^{\perp, a} (x,z) &=& g_5 \,  v^{a,n}(z) \hat V_{\hat \mu}^{a,n} (x)  \,   , \label{KKExpV}\\
\tilde \pi^a(x,z) &=& g_5 \, \tilde \pi^{a,n}(z) \hat{\pi}_V^{a,n}(x) \, , \label{KKExpTPi}\\
\tilde \phi^a(x,z) &=& g_5 \,  \tilde \phi^{a,n}(z) \hat{\pi}_V^{a,n}(x)
\, , \label{KKExpTPhi} \, ,
\eeqa
where a sum from $n=0$ to $n=\infty$ is implicit.  A similar decomposition holds for the 5-d fields in the axial sector:
\beqa
A_{\hat \mu}^{\perp, a} (x,z) &=& g_5 \,  a^{a,n}(z) \hat A_{\hat \mu}^{a,n} (x)  \, , \label{KKExpA} \\
 \pi^{a}(x,z) &=& g_5 \,  \pi^{a,n}(z) \hat \pi^{a,n}(x)  \, , \label{KKExpPi} \\
\phi^{a}(x,z) &=& g_5 \,  \phi^{a,n}(z) \hat \pi^{ a,n}(x) \,.  \label{KKExpPhi}
\eeqa
Using these expansions the actions in Eqs.~(\ref{SV2}) and (\ref{SA2}) factorize into $z$ integrals and $x$
integrals and we find $S^{(2)}_{V} = \int d^4 x {\cal L}_{V}$ and $S^{(2)}_{A} = \int d^4 x {\cal L}_{A}$,
with the vector and axial 4-d Lagrangians given by
\beqa
{\cal L}_{V} &=&  - \frac14 \Delta_V^{a, nm} \hat v_{\hat \mu \hat \nu}^{a,n} \hat v^{\hat \mu \hat \nu}_{a,m}
+\frac12 M_V^{a,nm} \hat V_{\hat \mu}^{a,n} \hat V^{\hat \mu}_{a,m}  \cr
&+& \frac12 \Delta_{\pi_V}^{a,nm} (\partial_{\hat \mu} \hat \pi_V^{a,n}) (\partial^{\hat \mu} \pi_V^{a,m})
- \frac12 M_{\pi_V}^{a,nm} \hat \pi_V^{a,n} \hat \pi_V^{a,m} \, , \cr
&& \label{LV}
\eeqa
\beqa
{\cal L}_{A} &=&  - \frac14  \Delta_A^{a, nm} \hat a_{\hat \mu \hat \nu}^{a,n} \hat a^{\hat \mu \hat \nu}_{a,m}
+\frac12  M_A^{a,nm} \hat A_{\hat \mu}^{a,n} \hat A^{\hat \mu}_{a,m} \cr
&+& \frac12  \Delta_{\pi}^{a,nm}(\partial_{\hat \mu} \hat \pi^{ a,n})(\partial^{\hat \mu} \hat \pi^{a,m})
-\frac12  M_{\pi}^{a,nm} \hat \pi^{a,n} \hat \pi^{a,m}
 \, , \cr
&& \label{LA}
\eeqa
with coefficients defined by the $z$ integrals
\beqa
\Delta_V^{a,nm} &=& \int \frac{dz}{z} v^{a,n}(z) v^{a,m}(z) , \cr
M_V^{a,nm} &=& \int \frac{dz}{z} \Big \{ [ \partial_z v^{a,n}(z) ] [ \partial_z v^{a,m}(z) ] \cr
&+& \beta^a_V(z) v^{a,n}(z) v^{a,m}(z) \Big \} , \cr
\Delta_{\pi_V}^{a,nm} &=& \int \frac{dz}{z} \Big \{ [ \partial_z \tilde \phi^{a,n} (z) ]  [ \partial_z \tilde \phi^{a,m} (z) ] \cr
&+& \beta^{a}_V(z) [ \tilde \pi^{a,n}(z) - \tilde \phi^{a,n}(z) ] [ \tilde \pi^{a,m}(z) - \tilde \phi^{a,m}(z) ]
\Big \} , \cr
M_{\pi_V}^{a,nm} &=& \int \frac{dz}{z} \beta_V^a(z) [ \partial_z \tilde \pi^{a,n} ] [ \partial_z \tilde \pi^{a,m} ],
\label{vectorcondition}
\eeqa
for the vector sector and
\beqa
\Delta_A^{a,nm} &=& \int \frac{dz}{z} a^{a,n}(z) a^{a,m}(z)  \, , \cr
M_A^{a ,nm} &=& \int \frac{dz}{z} \Big \{ [ \partial_z a^{a,n}(z) ] [ \partial_z a^{a ,m}(z) ] \cr
&+& \beta^{a}_A(z) a^{a,n}(z) a^{a ,m}(z) \Big \}  \, , \cr
\Delta_{\pi}^{a,nm} &=& \int \frac{dz}{z} \Big \{ [ \partial_z  \phi^{a,n} (z) ]  [ \partial_z  \phi^{a,m} (z) ]  \cr
&+& \beta^{a}_A(z) [  \pi^{a,n}(z) -  \phi^{a,n}(z) ] [ \pi^{a,m}(z) -  \phi^{a,m}(z) ]  \Big \} \, , \cr
M_{\pi}^{a,nm} &=& \int \frac{dz}{z} \beta_A^{a}(z) [ \partial_z  \pi^{a,n} ] [ \partial_z  \pi^{ a,m} ],
\label{axialcondition}
\eeqa
for the axial sector. Here we have defined
\beqa
\beta_V^a := \frac{g_5^2}{z^2}  (M_V^a)^2  \quad , \quad
\beta_A^{a} := \frac{g_5^2}{z^2} M_A^{aa} .
\eeqa

In order to obtain standard kinetic terms in Eqs.~(\ref{LV}) and (\ref{LA}) we impose the following
conditions:
\beqa
&&\Delta_V^{a, n m} = \Delta_{\pi_V}^{a, nm} =
\Delta_A^{a , nm} = \Delta_{\pi}^{a , n m} = \delta^{nm} \, , \cr
&& M_V^{a , nm} = m_{V^{a,n}}^2 \delta^{nm} \quad , \quad
M_{\pi_V}^{a, nm} = m_{\pi_V^{a,n}}^2 \delta^{nm}
\, , \cr
&& M_A^{a , nm} = m_{A^{a,n}}^2 \delta^{nm} \quad , \quad
M_{\pi}^{a, nm} = m_{\pi^{a,n}}^2 \delta^{nm}
\, . \cr
&&
\eeqa
The Lagrangians in Eqs.~(\ref{LV}) and (\ref{LA}) then reduce to
\beqa
{\cal L}_V &=&  - \frac14  \hat v_{\hat \mu \hat \nu}^{a,n} \hat v^{\hat \mu \hat \nu}_{a,n}
+\frac12 m_{V^{a,n}}^2 \hat V_{\hat \mu}^{a,n} \hat V^{\hat \mu}_{a,n}  \cr
&+& \frac12  (\partial_{\hat \mu} \hat \pi_V^{a,n}) (\partial^{\hat \mu} \pi_V^{a,n})
- \frac12 m_{\pi_V^{a,n}}^2 \hat \pi_V^{a,n} \hat \pi_V^{a,n}  \, ,  \label{LV2}
\eeqa
\beqa
{\cal L}_A &=&  - \frac14   \hat a_{\hat \mu \hat \nu}^{a,n} \hat a^{\hat \mu \hat \nu}_{a,n}
+\frac12  m_{A^{a,n}}^2 \hat A_{\hat \mu}^{a,n} \hat A^{\hat \mu}_{a,n} \cr
&+& \frac12  (\partial_{\hat \mu} \hat \pi^{a,n})(\partial^{\hat \mu} \hat \pi^{a,n})
-\frac12  m_{\pi^{a,n}}^2 \hat \pi^{a,n} \hat \pi^{a,n}  \, . \label{LA2}
\eeqa
The conditions for the $\Delta$ coefficients are normalization rules for the corresponding wave
 functions. The conditions for the masses are equivalent to the conditions for the $\Delta$ coefficients
 if we impose the following equations :
\beqa
&&   \left [ - \partial_z \left ( \frac{1}{z} \partial_z   \right )
+  \frac{1}{z} \beta_V^a(z) \right ]  v^{a,n}(z)  = \frac{m_{V^{a,n}}^2}{z}
v^{a,n}(z) \, , \cr
&& \frac{\beta_V^a(z)}{z} \left[\tilde \pi^{a,n}(z) - \tilde \phi^{a,n}(z) \right] =
- \partial_z \left [\frac{1}{z} \partial_z \tilde \phi^{a,n}(z) \right ] \, , \cr
&& \beta_V^a(z) \, \partial_z \tilde \pi^{a,n}(z) = m_{\pi_V^{a,n}}^2 \partial_z
\tilde \phi^{a,n}(z)
\, , \label{ModeEqsVector}
\eeqa
for the vector sector and
\beqa
&& \left [ - \partial_z \left ( \frac{1}{z} \partial_z   \right )
+  \frac{1}{z} \beta_A^{a}(z) \right ]  a^{a,n}(z) = \frac{m_{A^{ a,n}}^2}{z} a^{a,n}(z)
 \, , \cr
 && \frac{\beta_A^{ a}(z)}{z} \left[\pi^{a,n}(z) - \phi^{a,n}(z) \right] =
- \partial_z \left [\frac{1}{z} \partial_z \phi^{a,n}(z) \right ] \, , \cr
&& \beta_A^{a}(z) \, \partial_z \pi^{a,n}(z) = m_{\pi^{a,n}}^2 \partial_z
\phi^{a,n}(z) \, , \label{ModeEqsAxial}
\eeqa
for the axial sector.

We finish this section writing the $SU(4)$ pseudoscalar and vector  meson matrices $\hat \pi$ and $\hat V$
in terms of the charged states
\begin{widetext}
\beqa
\hat \pi &=& \hat \pi^a T^a = \frac{1}{\sqrt 2}
\left (\begin{matrix}
  \frac{\pi^0}{\sqrt{2}}  + \frac{\eta}{\sqrt{6}}  + \frac{\eta_c}{\sqrt{12}}  &  \pi^{+}  &  K^{+}  &  \bar D^0  \\
  \pi^{-}   & -\frac{\pi^0}{\sqrt{2}}  + \frac{\eta}{\sqrt{6}}  + \frac{\eta_c}{\sqrt{12}}   &  K^0  &  D^{-}  \\
   K^{-}  &  \bar K^0  & - \sqrt{\frac23} \eta + \frac{\eta_c}{\sqrt{12}}  &  D_s^{-}  \\
   D^0 &  D^{+}  &  D_s^{+}  & - \frac{3}{\sqrt{12}} \eta_c
 \end{matrix} \right ) , \label{pimatrix} \\[0.3true cm]
\hat V &=&  \hat V^a T^a  = \frac{1}{\sqrt 2}
\left ( \begin{matrix}
  \frac{\rho^0}{\sqrt{2}} + \frac{\omega'}{\sqrt{6}} + \frac{\psi}{\sqrt{12}}  &  \rho^{+}  &  K^{\star +}  &  \bar D^{\star 0}  \\
  \rho^{-}   & -\frac{\rho^0}{\sqrt{2}}  + \frac{\omega'}{\sqrt{6}}  + \frac{\psi}{\sqrt{12}}   &  K^{\star 0}  &  D^{\star -}  \\
   K^{\star -}  &  \bar K^{\star 0}  & - \sqrt{\frac23} \omega' + \frac{\psi}{\sqrt{12}}  &  D_s^{\star -}  \\
   D^{\star 0} &  D^{\star +}  &  D_s^{\star +}  & - \frac{3}{\sqrt{12}} \psi
 \end{matrix} \right ), \label{Vmatrix}
\eeqa
where in the last equation we have omitted the index $\mu$ for simplicity.
\end{widetext}

\section{Decay constants, CSB and FSB}

\label{Sec:Decayconstants}

As observed in \cite{Ballon-Bayona:2014oma}, the simplest method for extracting the leptonic decay
constants is to replace the fields in the r.h.s of the dual currents prescription,  Eqs.~(\ref{JV})-(\ref{JPi}),
by their  Kaluza-Klein expansions in Eqs.~(\ref{KKExpV})-(\ref{KKExpPhi}). We find:
\beqa
\langle J^{\hat \mu}_{V,a}  (x) \rangle &=&
\left [  \frac{1}{g_5 z} \partial_z v^{a,n}(z)  \right ]_{z=\epsilon} \hat V^{\hat \mu}_{a,n}(x)  \cr
&+&
\left [  \frac{1}{g_5 z} \partial_z \tilde \phi^{a,n}(z)  \right ]_{z=\epsilon} \partial^{\hat \mu} \hat \pi^{a,n}_V(x) \, ,
\label{JVExp}\\
\langle J^{\hat \mu}_{A,  a}  (x) \rangle &=&    \left [  \frac{1}{g_5 z }
\partial_z a^{ a ,n}(z)   \right ]_{z=\epsilon} \hat A^{\hat \mu}_{ a,n}(x) \cr
&+&  \left [ \frac{1}{g_5 z} \partial_z \phi^{ a , n}(z) \right ]_{z=\epsilon}
\partial^{\hat \mu} \hat \pi^{a ,n}(x) , \label{JAExpDiag} \\
\partial_{\hat \mu} \langle J^{\hat \mu}_{A, a} (x) \rangle &=& \langle J_{\Pi , a}  (x) \rangle \cr
&=&   -  \left [  \frac{ \beta^{a}_A (z) }{g_5 z}\partial_z \pi^{ a, n}(z)
\right ]_{z=\epsilon} \hat \pi^{a , n}(x) ,
\label{JPiExpDiag}
\eeqa
where a sum from $n=0$ to $n=\infty$ is implicit. In the expansions (\ref{JVExp})-(\ref{JPiExpDiag})
the 4-d fields $\hat V^{\hat \mu}_{a,n}(x) , A^{\hat \mu}_{a,n}(x) ,  \hat \pi^{a,n}_V(x) $ and $\hat \pi^{\bar a , n}(x)$ are on-shell. From these expansions we find the holographic prescription for leptonic decay
constants:
\beqa
g_{V^{a,n}} &=& \left [  \frac{1}{g_5 z} \partial_z v^{a,n}(z)  \right ]_{z=\epsilon} \, , \label{gV} \\
f_{\pi_V^{a,n}} &=& - \left [  \frac{1}{g_5 z} \partial_z \tilde \phi^{a,n}(z)  \right ]_{z=\epsilon}   \, , \label{fpiV} \\
g_{A^{a , n}} &=& \left [  \frac{1}{g_5 z }
\partial_z a^{a ,n}(z)   \right ]_{z=\epsilon} \, , \label{gA} \\
f_{\pi^{ a , n}} &=& - \left [ \frac{1}{g_5 z} \partial_z \phi^{ a , n}(z) \right ]_{z=\epsilon} \, . \label{fpi}
\eeqa
Taking the divergence of Eqs.~(\ref{JVExp}) and (\ref{JAExpDiag}) we find
\beqa
\partial_{\hat \mu} \langle J^{\hat \mu}_{V,a}  (x) \rangle =   f_{\pi_V^{a,n}} m_{\pi_V^{a,n}}^2 \hat \pi^{a,n}_V(x)  \, , \label{FSB}
\eeqa
and
\beqa
\partial_{\hat \mu} \langle J^{\hat \mu}_{A,a}  (x) \rangle =   f_{\pi^{a,n}} m_{\pi^{a,n}}^2 \hat \pi^{a,n}(x)
\, , \label{PCAC}
\eeqa
where a sum from $n=0$ to $n=\infty$ is implicit. Equation (\ref{PCAC}) is a generalization of the partially
conserved axial current relation (PCAC), which encodes the effect of chiral symmetry breaking (CSB) in the
current algebra. Equation (\ref{FSB}) encodes the effect of flavor symmetry breaking (FSB) in the vector current. Interestingly, the scalar mesons $\hat \pi^{a,n}_V(x)$ of FSB  and the pions  $\hat \pi^{a,n}(x)$ of CSB
appear in a similar way in Eqs.~(\ref{FSB}) and (\ref{PCAC}), respectively.

\section{Coupling Constants and Form Factors}
\label{sec:couplings}

\label{Sec:Couplings}

The three-point interactions are described by the 5-d action in Eq.~(\ref{S3}). After decomposing the fields
into their $(z,\mu)$ components and evaluating the metric in Eq.~(\ref{AdSMetric}), the action takes the
form
\beqa
S_3 = S_{VVV} + S_{VAA} + S_{V A  \pi} + S_{V  \pi   \pi} \, ,
\eeqa
where
\beqa
S_{VVV} &=& - \frac{1}{2g_5^2} f^{abc} \int d^4 x \int \frac{dz}{z} \Big [ v^{\hat \mu \hat \nu}_a V_{\hat \mu}^b V_{\hat \nu}^c \cr
&-& 2 v_{z \hat \mu}^a V_z^b V^{\hat \mu }_c  \Big ] \, , \label{SVVV}
\eeqa
\beqa
S_{VAA} &=& \frac{1}{2 g_5^2}  f^{abc} \int d^4 x \int \frac{dz}{z} \Big [ v^{\hat \mu \hat \nu}_a A_{\hat \mu}^b A_{\hat \nu}^c - 2 v_{z \hat \mu}^a A_z^b A_c^{\hat \mu} \cr
&-& 2 a^{\hat \mu \hat \nu}_a V_{\hat \mu}^b A_{\hat \nu}^c + 2 a_{z \hat \mu}^a (V_z^b A^{\hat \mu}_c
- V^{\hat \mu}_b A_z^c )\Big ] \, , \label{SVAA}
\eeqa
\beqa
S_{VA \, \pi} &=& \frac{1}{g_5^2} f^{abc} \int d^4 x \int \frac{dz}{z} \left [ \beta_V^b(z)  - \beta_A^{a} (z)  \right ]  \cr
&\times& \left [ - A_z^a V_z^b + A_{\hat \mu}^a V^{\hat \mu}_b\right ] \pi^c \, , \label{SVApi}
\eeqa
\beqa
S_{V  \pi   \pi} &=& \frac{1}{2g_5^2} f^{abc} \int d^4 x \int \frac{dz}{z} \left [ -  \beta_V^b(z) + 2 \beta_A^{a}(z)  \right ] \cr
&\times& \left [ - (\partial_z \pi^a) V_z^b + (\partial_{\hat \mu} \pi^a ) V^{\hat \mu}_b \right ] \pi^ c \, . \label{SVpipi}
\eeqa

Here we are interested in the following 4-d triple couplings
\beqa
S_{\hat V \hat V \hat V} =  g_{_{  \hat V^{a , \ell} \hat V^{b, m} \hat V^{c,n} }} \int d^4 x \, \hat V^{\hat \mu}_{a ,\ell} \, \hat  v_{\hat \mu \hat \nu}^{b, m} \,  V^{\hat \nu}_{c,n} \, , \label{4dVVV}
\eeqa
\beqa
S_{\hat V \hat \pi  \hat \pi} =  g_{_{\hat \pi^{a,\ell} \hat V^{b,m} \hat \pi^{c,n} }} \int d^4 x \, (\partial_{\hat \mu} \hat \pi^{a , \ell} ) \hat V^{\hat \mu}_{b,m} \hat \pi^{c,n} , \label{4dpiVpi}
\eeqa
where a sum over $a,b,c$ as well as $\ell, m, n$ is implicit.

Using (\ref{Vmu})-(\ref{AGauge})  in (\ref{SVVV}), (\ref{SVAA}) and (\ref{SVpipi}) as well as the KK expansions (\ref{KKExpV}), (\ref{KKExpPi}) and (\ref{KKExpPhi})  we find that
\beqa
 g_{_{\hat V^{a , \ell} \hat V^{b, m} \hat V^{c,n} }} &=& \frac{g_5}{2} f^{abc} \int \frac{dz}{z} v^{a,\ell} (z) v^{b,m}(z) v^{c,n}(z) \, , \cr
 &&  \label{gVVV}
\eeqa
\beqa
&&g_{_{\hat \pi^{ a,\ell} \hat V^{b,m} \hat \pi^{c,n} }} = \frac{g_5}{2} f^{a b c} \int \frac{dz}{z} \Big \{ 2 (\partial_z \phi^{ a , \ell}) v^{b,m} (\partial_z \phi^{ c ,n}) \cr
&+& \left [ - \beta_V^b (z) + 2 \beta_A^{ a }(z) \right ] (\pi^{a,\ell}- \phi^{a,\ell}) v^{b,m} ( \pi^{c,n} - \phi^{c,n}) \Big \} \,. \cr
&& \label{gpiVpi}
\eeqa

In order to compare our results with chiral theory models we rewrite the 3-point interactions
in Eqs.~(\ref{4dVVV}) and (\ref{4dpiVpi}) as
\beqa
S_{\hat V \hat V \hat V} =  2 f^{abc}   \bar g_{_{\hat V^{a , \ell} \hat V^{b, m} \hat V^{c,n} }}
\int d^4 x \, \hat V^{\hat \mu}_{a ,\ell} \,  (\partial_{\hat \mu} \hat  V_{\hat \nu}^{b, m}) \,  V^{\hat \nu}_{c,n} \, ,
\cr
&& \label{4dVVVv2}
\eeqa
\beqa
S_{\hat V \hat \pi  \hat \pi} &=& f^{abc} \bar g_{_{\hat \pi^{a,\ell} \hat V^{b,m} \hat \pi^{c,n} }}
\int d^4 x \,  \hat V^{\hat \mu}_{b,m}  \Big [ (\partial_{\hat \mu} \hat \pi^{a , \ell} ) \hat \pi^{c,n} \cr
&-& (\partial_{\hat \mu} \hat \pi^{c , n} ) \hat \pi^{a,\ell}  \Big ]\, ,  \label{4dpiVpiv2}
\eeqa
where
\beqa
 \bar g_{_{\hat V^{a , \ell} \hat V^{b, m} \hat V^{c,n} }} = \frac{g_5}{2}
 \int \frac{dz}{z} v^{a,\ell} (z) v^{b,m}(z) v^{c,n}(z) , \label{gVVVv2}
\eeqa
and
\beqa
\bar g_{_{\hat \pi^{a,\ell} \hat V^{b,m} \hat \pi^{c,n} }} &=& \frac{g_5}{8}  \int \frac{dz}{z}v^{b,m} \Big \{ 4 (\partial_z \phi^{a , \ell})  (\partial_z \phi^{c ,n}) \cr
&+& \left [ - 2 \beta_V^b (z) + 2 (\beta_A^{a}(z) +\beta_A^{c}(z)) \right ] \cr
&\times& (\pi^{a,\ell}- \phi^{a,\ell})  ( \pi^{c,n} - \phi^{c,n}) \Big \} \,. \label{gpiVpiv2}
\eeqa
To arrive at Eqs.~(\ref{4dVVVv2}) and (\ref{4dpiVpiv2}), we have integrated by parts the actions
in Eqs.~(\ref{4dVVV}) and (\ref{4dpiVpi}) and also used the transversality of the vector mesons
($ \partial_{\hat \mu} \hat V^{\hat \mu}_{a ,\ell}=0$). Note that the coupling in Eq.~(\ref{gpiVpiv2})
is symmetric when interchanging the pion flavor indices $a$ and $c$, as required by crossing symmetry.

We are interested in describing strong couplings involving  charmed mesons $D$ and $D^{*}$, strange mesons $K$  and $K^*$ as well as light mesons $\pi$ and $\rho$. Then in (\ref{4dVVVv2}) and (\ref{4dpiVpiv2})    we select only $a=(1,..,7)$ and $a=(9,..,12)$ for the pseudoscalar mesons $\pi^a$, whereas for the vector mesons we pick $a=(1..,7)$, $a=(9,..,12)$ and $a=(8,15)$. The other states are taken to zero.  The reason we include $a=(8,15)$ in the vectorial sector is because it contributes to the electromagnetic form factors of the $D$ and $D^{*}$ as shown below in this section.  Using also the results in (\ref{pimatrix})-(\ref{Vmatrix}) and evaluating the $SU(4)$ structure constants $f^{abc}$ we arrive at the effective Lagrangians
\begin{widetext}
\beqa
\hspace{-1cm}{\cal L}_{V \pi  \pi } &=& {\cal L}_{\pi D^{*} D} + {\cal L}_{\rho D D} + {\cal L}_{\omega' D D} + {\cal L}_{\psi D D} + {\cal L}_{\pi K^{*} K}  + {\cal L}_{\rho K K} +  {\cal L}_{\omega' K K} +  {\cal L}_{\rho \pi \pi},
\label{LVpipi}
\eeqa
\beqa
\hspace{-1cm}{\cal L}_{\text{VVV}} = {\cal L}_{\rho D^{*} D^{*}} +  {\cal L}_{\omega' D^{*} D^{*}} +  {\cal L}_{\psi D^{*} D^{*}}
+ {\cal L}_{\rho K^* K^*} + {\cal L}_{\omega' K^* K^*} +  {\cal L}_{\rho \rho \rho} \ ,
\label{LVVV}
\eeqa
where

\beqa
{\cal L}_{\pi D^{*} D} &=& i\sqrt{2} \, g_{\pi  D^{*} D} \left[D_{\mu }^{*+}
\left(\bar{D}^0\overleftrightarrow{\partial^\mu}\pi ^{-} \right)
+ D_{\mu }^{*-}\left( \pi^{+}\overleftrightarrow{\partial^\mu}  D^0 \right)
+ D_{\mu }^{*0}\left(D^{-}\overleftrightarrow{\partial^\mu} \pi ^{+} \right)
+ \bar{D}_{\mu }^{*0}\left( \pi^{-} \overleftrightarrow{\partial^\mu}   D^{+}  \right)\right] \nn \\[0.2true cm]
&+&  i g_{\pi D^{*} D }\left[ D_{\mu}^{*+}\left( \pi^0 \overleftrightarrow{\partial^\mu} D^{-} \right)
+ D_{\mu }^{*-}\left( D^{+} \overleftrightarrow{\partial^\mu} \pi^0 \right)
+ D_{\mu }^{*0}\left(\bar{D}^0 \overleftrightarrow{\partial^\mu} \pi^0 \right)
+ \bar{D}_{\mu}^{*0}\left(  \pi ^0 \overleftrightarrow{\partial^\mu} D^0 \right)\right], %\\[0.3true cm]
\eeqa

\beqa
{\cal L}_{\rho D D} &=& i\sqrt{2} \, g_{\rho D D }\left[
\rho_{\mu}^{+}\left(D^0 \overleftrightarrow{\partial^\mu}D^{-} \right) + \rho_{\mu}^{-}\left( D^{+}
\overleftrightarrow{\partial^\mu} \bar{D}^0 \right) \right]
+ ig_{\rho DD }\left[ \rho_{\mu}^0\left(D^-\overleftrightarrow{\partial^\mu}D^{+}\right)
+ \rho_{\mu}^0\left(D^0\overleftrightarrow{\partial^\mu} \bar{D}^0 \right)
\right], %\\[0.3true cm]
\eeqa

\beqa
{\cal L}_{\omega' D D} &=&
 \frac{i}{\sqrt{3}}  \, g_{\omega' D D}  \left [ \omega'_{\mu} \left ( D^{+} \overleftrightarrow{\partial^\mu} D^{-} \right )
 + \omega'_{\mu} \left ( D^0 \overleftrightarrow{\partial^\mu} \bar D^0 \right )  \right ], %\%\[0.3true cm]
\eeqa

\beqa
{\cal L}_{\psi D D} &=&
 i \sqrt{\frac83} \, g_{\psi D D}  \left [ \psi_{\mu} \left ( D^{+} \overleftrightarrow{\partial^\mu} D^{-} \right ) + \psi_{\mu} \left ( D^0 \overleftrightarrow{\partial^\mu} \bar D^0 \right ) \right ], %\\[0.3true cm]
\eeqa

\beqa
{\cal L}_{\rho D^{*} D^{*}} &=& i\sqrt{2} \, g_{\rho D^{*} D^{*}}\Bigl [ D_{\mu }^{*+}
\left (  \rho^{-}_{\nu} \overleftrightarrow{\partial^\mu}  \bar{D}^{\nu}_{*0} \right )
+  D_{\mu }^{*-} \left ( D_{\nu }^{*0} \overleftrightarrow{\partial^\mu} \rho_{+}^{\nu } \right )
+  D_{\mu}^{*0} \left ( \rho^{+}_{\nu }  \overleftrightarrow{\partial^\mu} D^{\nu }_{*-}  \right )
+ \bar{D}_{\mu}^{*0}
\left ( D_{\nu }^{*+} \overleftrightarrow{\partial^\mu} \rho_{-}^{\nu} \right )
\nn \\[0.2true cm]
&+&  \rho_{\mu}^{+} \left ( D^{*-}_{\nu}    \overleftrightarrow{\partial^\mu}  D^{\nu}_{*0} \right )
+ \rho_{\mu}^{-} \left ( \bar{D}_{\nu}^{*0}  \overleftrightarrow{\partial^\mu} D_{*+}^{\nu} \right )    \Big ]
+ i g_{\rho D^{*}D^{*}}\Big [  D_{\mu}^{*+} \left ( D_{\nu}^{*-} \overleftrightarrow{\partial^\mu} \rho_0^{\nu} \right )  + D_{\mu}^{*-} \left ( \rho^0_{\nu} \overleftrightarrow{\partial^\mu}  D^{\nu }_{*+}  \right )
\nn \\[0.2true cm]
&+&  D_{\mu}^{*0} \left ( \rho^0_{\nu} \overleftrightarrow{\partial^\mu} \bar{D}^{\nu}_{*0}  \right )
+ \bar{D}_{\mu}^{*0} \left ( D_{\nu }^{*0} \overleftrightarrow{\partial^\mu}\rho _0^{\nu} \right )
+  \rho_{\mu}^0 \left ( D_{\nu}^{*+} \overleftrightarrow{\partial^\mu} D_{*-}^{\nu} \right )
+   \rho_{\mu}^0 \left ( \bar{D}_{\nu}^{*0} \overleftrightarrow{\partial^\mu}  D_{*0}^{\nu} \right )
 \Big ], %\\[0.3true cm]
\eeqa

\beqa
{\cal L}_{\omega' D^{*} D^{*}} &=& - \frac{i}{\sqrt{3}} g_{\omega' D^{*} D^{*}} \Big [  D_{\mu}^{*+} \left ( D_{\nu}^{*-} \overleftrightarrow{\partial^\mu} \omega'^{\nu}  \right ) + D_{\mu}^{*-} \left ( \omega'^{\nu}  \overleftrightarrow{\partial^\mu}  D_{\nu}^{*+} \right ) + D_{\mu}^{*0} \left ( \bar D_{\nu}^{*0} \overleftrightarrow{\partial^\mu} \omega'^{\nu} \right )
 + \bar D_{\mu}^{*0} \left ( \omega'^{\nu}  \overleftrightarrow{\partial^\mu}   D_{\nu}^{*0}\right )
\nn \\
&+& \omega'_{\mu} \left (D_{\nu}^{*+} \overleftrightarrow{\partial^\mu} D_{*-}^{\nu}  \right )
+ \omega'_{\mu} \left (D_{\nu}^{*0} \overleftrightarrow{\partial^\mu} \bar D_{*0}^{\nu}  \right )
  \Big ],
\eeqa

\beqa
{\cal L}_{\psi D^{*} D^{*}} &=& - i \sqrt{\frac83} g_{\psi D^{*} D^{*}} \Big [  D_{\mu}^{*+} \left ( D_{\nu}^{*-} \overleftrightarrow{\partial^\mu} \psi^{\nu}  \right ) + D_{\mu}^{*-} \left ( \psi^{\nu} \overleftrightarrow{\partial^\mu} D_{\nu}^{*+}   \right )
+  D_{\mu}^{*0} \left ( \bar D_{\nu}^{*0} \overleftrightarrow{\partial^\mu} \psi^{\nu} \right )
+ \bar D_{\mu}^{*0} \left ( \psi^{\nu}  \overleftrightarrow{\partial^\mu} D_{\nu}^{*0}  \right )
\nn \\
&+& \psi_{\mu} \left (D_{\nu}^{*+} \overleftrightarrow{\partial^\mu} D_{*-}^{\nu}  \right )
+ \psi_{\mu} \left (D_{\nu}^{*0} \overleftrightarrow{\partial^\mu} \bar D_{*0}^{\nu}  \right )
  \Big ],
\eeqa

\beqa
{\cal L}_{\pi K^* K} &=& i \sqrt{2} \, g_{\pi K^* K} \Big [ K_{\mu}^{*+} \left ( \pi^{-} \overleftrightarrow{\partial^\mu} \bar K^0 \right ) + K_{\mu}^{*-} \left ( K^0 \overleftrightarrow{\partial^\mu} \pi^{+} \right )
+ K_{\mu}^{*0} \left ( \pi^{+} \overleftrightarrow{\partial^\mu} K^{-} \right )
+ \bar K_{\mu}^{*0}  \left ( K^{+} \overleftrightarrow{\partial^\mu} \pi^{-} \right ) \Big ] \nn \\[0.2true cm]
&+& i   g_{\pi K^* K} \Big [ K_{\mu}^{*+} \left ( \pi^0 \overleftrightarrow{\partial^\mu} K^{-}\right )
+ K_{\mu}^{*-} \left ( K^{+} \overleftrightarrow{\partial^\mu}  \pi^0 \right )
+ K_{\mu}^{*0} \left ( \bar K^0\overleftrightarrow{\partial^\mu} \pi^0 \right )
+ \bar K_{\mu}^{*0} \left ( \pi^0 \overleftrightarrow{\partial^\mu} K^0 \right ) \Big ] ,
\eeqa

\beqa
{\cal L}_{\rho K K} &=& i \sqrt{2} \, g_{\rho K K} \Big [ \rho_{\mu}^{+} \left ( K^{-} \overleftrightarrow{\partial^\mu} K^0 \right ) + \rho_{\mu}^{-} \left ( \bar K^0 \overleftrightarrow{\partial^\mu} K^{+} \right ) \Big ]
+ i g_{\rho K K} \Big [ \rho_{\mu}^0 \left ( K^{-} \overleftrightarrow{\partial^\mu} K^{+} \right ) + \rho_{\mu}^0 \left ( K^0 \overleftrightarrow{\partial^\mu} \bar K^0 \right ) \Big ],
\eeqa

\beqa
{\cal L}_{\omega' K K} &=& i \sqrt{3} \, g_{\omega' K K} \Big [ \omega'_{\mu} \left ( K^{-} \overleftrightarrow{\partial^\mu} K^{+} \right )
+\omega'_{\mu} \left ( \bar K^0 \overleftrightarrow{\partial^\mu} K^0 \right ) \Big ] ,
%\\[0.3true cm]
\eeqa

\beqa
{\cal L}_{\rho K^{*} K^{*}} &=& i \sqrt{2} g_{\rho K^{*} K^{*}} \Big [ K_{\mu}^{*+} \left ( \bar K_{\nu}^{*0} \overleftrightarrow{\partial^\mu} \rho_{-}^{\nu} \right ) + K_{\mu}^{*-} \left ( \rho_{\nu}^{+} \overleftrightarrow{\partial^\mu} K_{\nu}^{*0} \right )
+ K_{\mu}^{*0} \left ( K_{\nu}^{*-} \overleftrightarrow{\partial^\mu}\rho_{+}^{\nu} \right )
+ \bar K_{\mu}^{*0} \left ( \rho_{\nu}^{-} \overleftrightarrow{\partial^\mu}  K_{*+}^{\nu} \right )
\nn \\[0.2true cm]
&+& \rho_{\mu}^{+} \left ( K_{\nu}^{*0} \overleftrightarrow{\partial^\mu} K_{*-}^{\nu} \right )
+ \rho_{\mu}^{-} \left ( K_{\nu}^{*+} \overleftrightarrow{\partial^\mu} \bar K_{*0}^{\nu} \right ) \Big ]
+ i  g_{\rho K^{*} K^{*}}  \Big [ K_{\mu}^{*+} \left ( K_{\nu}^{*-} \overleftrightarrow{\partial^\mu} \rho_0^{\nu} \right ) + K_{\mu}^{*-} \left ( \rho_{\nu}^0 \overleftrightarrow{\partial^\mu} K_{*+}^{\nu} \right ) \nn \\[0.2true cm]
&+& K_{\mu}^{*0} \left ( \rho_{\nu}^0 \overleftrightarrow{\partial^\mu} \bar K_{*0}^{\nu} \right ) + \bar K_{\mu}^{*0} \left ( K_{\nu}^{*0} \overleftrightarrow{\partial^\mu} \rho_0^{\nu} \right )
+\rho_{\mu}^0 \left ( K_{\nu}^{*+} \overleftrightarrow{\partial^\mu} K_{*-}^{\nu} \right ) + \rho_{\mu}^0 \left ( \bar K_{\nu}^{*0} \overleftrightarrow{\partial^\mu} K_{*0}^{\nu} \right ) \Big ] ,
%\\[0.3true cm]
\eeqa

\beqa
{\cal L}_{\omega' K^* K^*} &=& i \sqrt{3} \, g_{\omega' K^* K^*} \Big [ K_{\mu}^{*+} \left ( K_{\nu}^{*-} \overleftrightarrow{\partial^\mu} \omega'^{\nu} \right )  + K_{\mu}^{*-} \left ( \omega'_{\nu} \overleftrightarrow{\partial^\mu} K_{*+}^{\nu} \right ) + K_{\mu}^{*0} \left ( \bar K_{\nu}^{*0} \overleftrightarrow{\partial^\mu} \omega'^{\nu} \right ) + \bar K_{\mu}^{*0} \left (  \omega'_{\nu} \overleftrightarrow{\partial^\mu} K_{*0}^{\nu} \right )
\nn \\[0.2true cm]
&+&  \omega'_{\mu} \left ( K_{\nu}^{*+}  \overleftrightarrow{\partial^\mu} K_{*-}^{\nu} \right )
+ \omega'_{\mu} \left ( K_{\nu}^{*0} \overleftrightarrow{\partial^\mu} \bar K_{*0}^{\nu} \right )   \Big ],
\eeqa

\beqa
{\cal L}_{\rho \pi \pi} &=& i g_{\rho \pi \pi }\Big [\rho_{\mu}^{+}\left(\pi^0 \overleftrightarrow{\partial^\mu}\pi^{-} \right)
+\rho_{\mu }^{-}\left(\pi^{+} \overleftrightarrow{\partial^\mu} \pi^0 \right)
+ \rho_{\mu}^0\left(\pi^{-} \overleftrightarrow{\partial^\mu} \pi ^{+}  \right)
\Big ], \\[0.3true cm]
{\cal L}_{\rho \rho \rho} &=& i g_{\rho \rho \rho } \Big[ \rho_{\mu}^{+} \left ( \rho_{\nu}^{-} \overleftrightarrow{\partial^\mu} \rho_0^{\nu} \right )  + \rho_{\mu}^{-}
\left ( \rho_{\nu}^0 \overleftrightarrow{\partial^\mu} \rho_{+}^{\nu} \right )
+ \rho_{\mu}^0 \left ( \rho _{\nu}^{+} \overleftrightarrow{\partial^\mu} \rho_{-}^{\nu} \right )
 \Big ] .
\eeqa
\end{widetext}
In the above, the couplings are given by
\beqa
g_{_{\pi D^{*} D}} &=& \bar g_{_{\hat \pi^{a} \hat V^{b}  \hat \pi^{c}}} \ , \quad a=(1,2,3) \ , \ (b,c)=(9,..,12) \ , \cr
g_{_{\rho D D}} &=& \bar g_{_{\hat \pi^{a} \hat V^{b}  \hat \pi^{c}}} \ , \quad (a,c)=(9,..,12) \ , \  b =(1,2,3) \ , \cr
g_{_{\omega' D D} } &=&  \bar g_{_{\hat \pi^{a} \hat V^{b}  \hat \pi^{c}}} \ , \quad (a,c)=(9,..,12) \ , \  b =8 \ , \cr
g_{_{\psi D D} } &=&  \bar g_{_{\hat \pi^{a} \hat V^{b}  \hat \pi^{c}}} \ , \quad (a,c)=(9,..,12) \ , \  b =15 \ , \cr
g_{_{\rho D^{*} D^{*}}} &=& \bar g_{_{\hat V^a \hat V^b \hat V^c}} \ , \quad  a=(1,2,3) \  , \  (b,c)=(9,..,12) \ ,  \cr
g_{_{\omega' D^{*} D^{*}}} &=& \bar g_{_{\hat V^a \hat V^b \hat V^c}} \ , \quad  a=8 \  , \  (b,c)=(9,..,12) \ , \cr
g_{_{\psi D^{*} D^{*}}} &=& \bar g_{_{\hat V^a \hat V^b \hat V^c}} \ , \quad  a=15 \  , \  (b,c)=(9,..,12) \ , \cr
g_{_{\pi K^{*} K}} &=& \bar g_{_{\hat \pi^{a} \hat V^{b}  \hat \pi^{c}}} \ , \quad a=(1,2,3) \ , \ (b,c)=(4,..,7) \ , \cr
g_{_{\rho K K}} &=& \bar g_{_{\hat \pi^{a} \hat V^{b}  \hat \pi^{c}}} \ , \quad (a,c)=(4,..,7) \ , \  b =(1,2,3) \ , \cr
g_{_{\omega' K K} } &=&  \bar g_{_{\hat \pi^{a} \hat V^{b}  \hat \pi^{c}}} \ , \quad (a,c)=(4,..,7) \ , \  b =8 \ , \cr
g_{_{\rho K^{*} K^{*}}} &=& \bar g_{_{\hat V^a \hat V^b \hat V^c}} \ , \quad  a=(1,2,3) \  , \  (b,c)=(4,..,7) \ ,  \cr
g_{_{\omega' K^{*} K^{*}}} &=& \bar g_{_{\hat V^a \hat V^b \hat V^c}} \ , \quad  a=8 \  , \  (b,c)=(4,..,7) \ , \cr
g_{_{\rho \pi \pi}} &=& 2 \bar g_{_{\hat \pi^{a} \hat V^{b}  \hat \pi^{c}}} \ , \quad  (a,b,c) =(1,2,3) \ , \cr
g_{_{\rho \rho \rho}} &=&  2 \bar g_{_{\hat V^a \hat V^b  \hat V^c}}\ , \quad  (a,b,c) =(1,2,3) \ . \label{RealCouplings}
\eeqa
We have used the double arrow derivative  $f\overleftrightarrow{\partial^\mu}g:= f (\partial^\mu g)
- (\partial^\mu f) g $ and for simplicity we have omitted the indices $\ell , m, n$ that distinguish the fundamental states from the corresponding resonances. The Lagrangians in Eqs.~(\ref{LVVV}) and (\ref{LVpipi}) are typically used in phenomenology
of charmed mesons{\textemdash}see e.g. Ref.\cite{Lin00a}.

In the limit where the quark masses and condensates are equal, flavor symmetry is recovered and the
couplings satisfy the relations
\beqa
g_{\pi D^{*} D} = g_{\rho D D } = g_{\omega' D D} = g_{\psi D D} = \cr
g_{\pi K^* K} = g_{\rho K K} = g_{\omega' K K}  = \frac12 g_{\rho \pi \pi} =: \frac{g}{4} \,,
\label{su4relation1}
\eeqa
\beqa
g_{\rho D^{*} D^{*}} = g_{\omega' D^{*} D^{*}} =  g_{\psi D^{*} D^{*}} = \cr
g_{\rho K^* K^*} = g_{\omega' K^* K^*} =   \frac12 g_{\rho \rho \rho}  := \frac{\tilde g}{4} \,.
\label{su4relation2}
\eeqa
In this case all the couplings can be obtained from the interaction terms $ i g {\rm Tr} (\partial^{\mu} \pi [ \pi, V_{\mu}])$ and $ i \tilde g {\rm Tr} (\partial^{\mu} V^{\nu} [ V_{\mu} , V_{\nu} ] )${\textemdash}see e.g.
Ref.~\cite{Lin:1999ve}.

\subsection{Electromagnetic form factors}

The effective Lagrangian (\ref{LVpipi}) describes the interaction between a vector meson and two pseudoscalar mesons.
If the vector meson is off-shell and the pseudoscalar mesons are on-shell we can use (\ref{LVpipi}) to investigate the electromagnetic (EM) form factors of pseudoscalar mesons. Similarly, using the effective Lagrangian (\ref{LVVV}) and taking one of the vector mesons off-shell we can investigate the EM form factors of vector mesons.
The (elastic) EM form factors of pseudoscalar mesons appear in the decomposition of the EM current as
\beqa
\left\langle \pi^{a} (p+q) \vert J^{\mu}_{EM} (0) \vert \pi^{a} (p) \right\rangle &=& (2p + q)^{\mu} F_{\pi^{a}}(q^{2}).
\eeqa
Similarly, the (elastic) EM form factors of vector mesons appear as the Lorentz scalars in the EM current decomposition \cite{Grigoryan:2007vg}
\begin{align}
&\left\langle V^a (p+q),\epsilon' \vert J^{\mu}_{EM} (0) \vert V^a (p),\epsilon \right\rangle \cr
&= -(\epsilon' \cdot \epsilon)
(2p + q)^{\mu} F_{V^a}^1(q^{2}) \cr
&+ [\epsilon'^{\mu}(\epsilon \cdot q)-\epsilon^{\mu}(\epsilon' \cdot q)]
[F_{V^a}^1(q^{2}) + F_{V^a}^2(q^{2})] \cr
&+ \frac{1}{M_{V^{a}}^{2}}(q \cdot \epsilon')(q \cdot \epsilon)(2p + q)^{\mu} F_{V^a}^3(q^{2})\, . \label{VectorVertex1}
\end{align}
Linear combinations of the form factors in (\ref{VectorVertex1}) define the so called electric, magnetic and quadrupole form factors:
\beqa
F_{V^a}^{E} &=& F_{V^a}^1 + \frac{q^2}{6 M_{V^{a}}^{2}}\Big[F_{V^a}^2 - \Big(1-\frac{q^2}{4M_{V^{a}}^{2}}\Big) F_{V^a}^3 \Big] \, , \cr
F_{V^a}^{M} &=& F_{V^a}^1 + F_{V^a}^2 \, , \cr
F_{V^a}^{Q} &=& - F_{V^a}^2 + \Big(1-\frac{q^2}{4M_{V^{a}}^{2}}\Big) F_{V^a}^3 \,. \label{EMQformfactors}
\eeqa
In the absence of baryonic number the EM current is obtained from a linear combination of flavor currents, i.e.
\beqa
J^{\mu}_{EM} (x) = \sum_{a=(3,8,15)} c_a J^{\mu}_a (x)\, , \label{EMCurrentDecomp}
\eeqa
where the coefficients $c_a$ can vary depending on the quarks that will be considered in the EM current.
When considering the EM form factors of the heavy-light $D$ and $D^*$ charmed mesons the strange quark does not participate in the process and we can define the EM
current as
\beqa
J^{\mu}_{EM} = \frac23 \bar u \gamma^{\mu} u - \frac13 \bar d \gamma^{\mu} d + \frac23 \bar c \gamma^{\mu} c .
\label{EMcurrent1}
\eeqa
Then the EM current can be decomposed as (\ref{EMCurrentDecomp}) with coefficients  $c_3=1$, $c_8=7/(3\sqrt{3})$ and $c_{15}=- 8/(3\sqrt{6})$, up to the strangeness current which do not contribute when evaluating the current at the external states. On the other hand, when evaluating the EM form factors for the strange $K$ and $K^*$ as well as the light mesons $\pi$ and $\rho$ we define the EM current as
\beqa
J^{\mu}_{EM} = \frac23 \bar u \gamma^{\mu} u - \frac13 \bar d \gamma^{\mu} d - \frac13 \bar s \gamma^{\mu} s ,
\label{EMcurrent2}
\eeqa
which admits the decomposition (\ref{EMCurrentDecomp}) for the coefficients $c_3=1$, $c_8 = 1/\sqrt{3}$ and $c_{15}=0$.

As explained in the previous section, each flavor current admits a decomposition in terms of vector mesons.
This implies from (\ref{EMCurrentDecomp}) that the photon decays into $\rho^{0,n}$, \, $\omega'^n$ and $\psi^n$ mesons. This is is a holographic realization of generalized vector meson dominance (GVMD)~\cite{Sakurai:1972wk}, in that also the resonances are included and not only the fundamental states as VMD.

For the pion and $\rho$ meson only the states $\rho^{0,n}$ contribute to the EM form factors. In the case of strange mesons $K$ and $K^*$ the states $\rho^{0,n}$ and $\omega'^n$ contribute to the EM form factors whereas in the case of the charmed $D$ and $D^*$ mesons we have contributions from $\rho^{0,n}$, $\omega'^n$ and $\psi^n$. In our model it turns out that the states $\omega'^n$ and $\psi^n$ are identical to the states  $\rho^{0,n}$ as well as their couplings to external states. Although this implies an unrealistic spectrum for those mesons, its contribution to the EM form factors is not only required by consistency but also leads to reasonable results consistent either with experimental data or lattice QCD data, as we will show below.

Using the Feynman rules for the vector meson propagator in  (\ref{LV2}) and the triple vertex (\ref{LVpipi}) as well as the EM current decomposition (\ref{EMCurrentDecomp}), with the appropriate coefficients, we extract the (elastic) EM form factors for the pion, kaon  and $D$ meson :
\beqa
F_{\pi}(Q^2)&=& \sum_{n} \frac{g_{\rho^n} g_{\rho^n \pi \pi}}{m_{\rho^n}^2 + Q^2} \, ,
\label{PionFormFactor} \\
F_{K}(Q^2) &=& \sum_{n} \Big [\frac{g_{\rho^n} g_{\rho^n K K}}{m_{\rho^n}^2 + Q^2} +  \frac{g_{\omega'^n} g_{\omega'^n K K}}{m_{\omega'^n}^2 + Q^2} \Big ]\cr
&=& 2 \sum_{n} \frac{g_{\rho^n} g_{\rho^n K K}}{m_{\rho^n}^2 + Q^2} \, ,
\label{KFormFactor} \\
F_{D}(Q^2) &=& \sum_{n} \Big [\frac{g_{\rho^n} g_{\rho^n D D}}{m_{\rho^n}^2 + Q^2} - \frac79  \frac{g_{\omega'^n} g_{\omega'^n D D}}{m_{\omega'^n}^2 + Q^2} \cr
&+&   \frac{16}{9}  \frac{g_{\psi^n} g_{\psi^n D D}}{m_{\psi^n}^2 + Q^2} \Big ] = 2 \sum_{n} \frac{g_{\rho^n} g_{\rho^n D D}}{m_{\rho^n}^2 + Q^2} \, ,
\label{DFormFactor}
\eeqa
where $Q^2 = - q^2$. The second equalities in (\ref{KFormFactor}) and (\ref{DFormFactor}) come from the identification of the states $\omega'^n$ and $\psi^n$ with the states $\rho^{0,n}$. Similarly for the vector sector, we use the Feynman rules associated with the triple vertex (\ref{LVVV}) and the vector meson propagator in  (\ref{LV2}) to extract the (elastic) EM form factors for the $\rho$ meson, $K^*$ meson and $D^{*}$ meson :
\beqa
F_{\rho}^1 = F_{\rho}^2 = F_{\rho}(Q^2) &=& \sum_{n} \frac{g_{\rho^n} g_{\rho^n \rho \rho}}{m_{\rho^n}^2 + Q^2} \, ,
\label{rhoFormFactor} \\
F_{K^{*}}^1 = F_{K^{*}}^2 = F_{K^{*}}(Q^2) &=& 2 \sum_{n} \frac{g_{\rho^n} g_{\rho^n K^{*} K^{*}}}{m_{\rho^n}^2 + Q^2} \, ,
\label{K*FormFactor} \\
F_{D^{*}}^1 = F_{D^{*}}^2 = F_{D^{*}}(Q^2) &=& 2 \sum_{n} \frac{g_{\rho^n} g_{\rho^n D^{*} D^{*}}}{m_{\rho^n}^2 + Q^2} \, ,
\label{D*FormFactor} \\
F_{\rho}^3 =  F_{K^{*}}^3  = F_{D^{*}}^3 &=& 0 \,.
\eeqa
The electric, magnetic and quadrupole form
factors are obtained using (\ref{EMQformfactors}).

\subsection{Low and high $Q^2$}

At low $Q^2$ the EM form factor of a pseudoscalar meson can be expanded as
\begin{eqnarray}
F_{\pi^{a}}(Q^{2})= 1 -  \frac{1}{6} \langle r_{\pi^{a}}^{2} \rangle Q^{2} + ... ,
\label{PSFormFactorLowQ2}
\end{eqnarray}
where the second term defines the charge radius. A similar expression follows for the vector mesons.
Notice that we have used the relation $F_{\pi^a}(0)=1$, which is due to charge conservation.
In fact, the relation $F_{\pi^a}(0)=1$ follows nicely from the sum
rules
\beqa
\sum_{n} \frac{g_{\rho^n} g_{\rho^n \pi \pi}}{m_{\rho^n}^2}  &=& 2 \sum_{n} \frac{g_{\rho^n} g_{\rho^n K K}}{m_{\rho^n}^2} \cr
&=& 2 \sum_{n} \frac{g_{\rho^n} g_{\rho^n D D}}{m_{\rho^n}^2} = 1 \, .
\label{sumrule1}
\eeqa
These sum rules can be proven using the equation and completeness relation of vector mesons as well
as the normalization of the external states.  For the vector mesons the electric radius is obtained from the electric form factor as
\beqa
\left\langle r_{V^{a}}^2 \right\rangle = -6 \left. \frac{d F_{V^a}^E(Q^2)}{dQ^2}\right\vert _{Q^2=0} \, .
\label{VMchargeradius}
\eeqa
The magnetic $\mu$ and quadrupole $D$ moments of the vector
mesons in our model take the canonical values
\beqa
\mu &=& F_{V^a}^{M}(0) =  2 \, , \cr
D   &=& \frac{1}{M_{V^{a}}^{2}}F_{V^a}^{Q}(0)= - \frac{1}{M_{V^{a}}^{2}} \,,
\eeqa
where we have used the relation $F_{V^a}(0)=1$ which follows from the sum rules
\beqa
\sum_{n} \frac{g_{\rho^n} g_{\rho^n \rho \rho}}{m_{\rho^n}^2} &=& 2 \sum_{n} \frac{g_{\rho^n} g_{\rho^n K^{*} K^{*}}}{m_{\rho^n}^2} \cr
&=& 2 \sum_{n} \frac{g_{\rho^n} g_{\rho^n D^{*} D^{*}}}{m_{\rho^n}^2}=1 \,. \label{sumrule2}
\eeqa
Again these sum rules follow from the equation and completeness of the $\rho^n$ states and the normalization
of the external states.  In fact, the sum rules Eqs.~(\ref{sumrule1}) and (\ref{sumrule2}) are universal in
bottom up and top-down holographic models for QCD. A discussion of these sum rules in the top-down approach
can be found in Ref.~\cite{Bayona:2010bg}.

In the regime of large $Q^2$, the EM form factors of pseudoscalar mesons can be expanded as
\beqa
F_{\pi}(Q^{2})&=& \frac{1}{Q^2} \sum_{n=0}^{\infty} g_{\rho^{n}}g_{\rho^n \pi \pi }
\Big [ 1 - \frac{m_{\rho^{n}}^{2}}{Q^2} + \dots \Big ]
\, , \cr
F_{K}(Q^{2})&=& \frac{2}{Q^2} \sum_{n=0}^{\infty} g_{\rho^{n}}g_{\rho^n K K }
\Big [ 1 - \frac{m_{\rho^{n}}^{2}}{Q^2} + \dots \Big ] \, \cr
F_{D}(Q^{2})&=& \frac{2}{Q^2} \sum_{n=0}^{\infty} g_{\rho^{n}}g_{\rho^n D D }
\Big [ 1 - \frac{m_{\rho^{n}}^{2}}{Q^2} + \dots \Big ] .\label{PSFormFactorHighQ2}
\eeqa
A similar expansion holds for the EM form factors of vector mesons
\begin{align}
F_{\rho}(Q^{2})&= \frac{1}{Q^2} \sum_{n=0}^{\infty} g_{\rho^{n}}g_{\rho^n \rho \rho }
\Big [ 1 - \frac{m_{\rho^{n}}^{2}}{Q^2} + \dots \Big ]
\, , \cr
F_{K^{*}}(Q^{2})&= \frac{2}{Q^2} \sum_{n=0}^{\infty} g_{\rho^{n}}g_{\rho^n K^{*} K^{*} }
\Big [ 1 - \frac{m_{\rho^{n}}^{2}}{Q^2} + \dots \Big ]
\, , \cr
F_{D^{*}}(Q^{2})&= \frac{2}{Q^2} \sum_{n=0}^{\infty} g_{\rho^{n}}g_{\rho^n D^{*} D^{*} }
\Big [ 1 - \frac{m_{\rho^{n}}^{2}}{Q^2} + \dots \Big ] .
\label{VMFormFactorHighQ2}
\end{align}

In the next section we present our predictions for the couplings and form factors involving the pions, kaons, $\rho$ mesons, $K^*$ mesons as well as the charmed mesons $D$ and $D^{*}$. For the $D$ and $D^*$ EM form factors we  compare our results against data from Lattice QCD. Using
(\ref{PSFormFactorLowQ2}) and (\ref{VMchargeradius}) we will also extract the charge radii of all those mesons and compare against experimental data or lattice QCD data. Last but not least, the high-$Q^2$ behavior of the form factors in Eqs.~(\ref{PSFormFactorHighQ2})-(\ref{VMFormFactorHighQ2}) will be checked and compared with perturbative QCD calculations.

%%%%%%%%%%%%%%%%%%%%%%%%%%%%%%%%%%%%%%%%%%%%%%%%%%%%%%%%%%%%%%%%%%%%%%%%%%%%%%%%%%%%%%%%%%%
\section{Results and comparison with Lattice QCD}

\label{Sec:Numerics}

In this section we present our numerical results for the spectrum, decay constants, coupling constants and EM form factors  involving the charmed mesons. It is convenient to define unnormalized wave functions for the scalar mesons ($\tilde \phi_U^{a,n}$ and $\tilde \pi_U^{a,n}$), pseudoscalar mesons ($\phi_U^{a,n}$ and $\pi_U^{a,n}$), vector mesons ($v_U^{a,n}$) and axial vector mesons ($a_U^{a,n}$) so that the first coefficient in the near boundary expansion is fixed arbitrarily (due to the linearity of the differential equations). Eqs.~(\ref{ModeEqsVector}) and (\ref{ModeEqsAxial}) dictate the near boundary behavior of  the unnormalized wave functions:
\beqa
\tilde \phi^{a,n}_U (z) &=&  -z^{2} + \dots \quad  , \quad
\tilde \pi^{a,n}_U (z) =  -\frac{m_{\pi_{V}^{a,n}}^{2}}{\beta_{V}^{a}(0)}z^{2} + \dots \,, \cr
\phi^{a,n}_U (z) &=&  -z^{2} + \dots  \quad  , \quad
 \pi^{a,n}_U (z) =  -\frac{m_{\pi^{a,n}}^{2}}{\beta_{A}^{a}(0)}z^{2} + \dots \, , \cr
v_U^{a,n}(z) &=& z^2 + \dots \quad  , \quad
a_U^{a,n}(z) = z^2 + \dots \, . \label{Asymptotics}
\eeqa
In Eq.~(\ref{Asymptotics}), the first coefficients were fixed to $1$ or $-1$ to guarantee a positive
sign for the decay constants in Eqs.~(\ref{gV})-(\ref{fpi}) and positive normalization constants.
The normalized wave functions take the form
\beqa
\hspace{-0.25cm}
\tilde \phi^{a,n}(z) &=& N_{\pi_V^{a,n}} \tilde \phi_U^{a,n}(z)
\, \,  , \, \,
\tilde \pi^{a,n}(z) = N_{\pi_V^{a,n}} \tilde \pi_U^{a,n}(z) \, , \cr
\hspace{-0.25cm}\phi^{a,n}(z) &=& N_{\pi^{a,n}} \phi_U^{a,n}(z)
\, \,  , \, \,
\pi^{a,n}(z) = N_{\pi^{a,n}} \pi_U^{a,n}(z) \, , \cr
\hspace{-0.25cm}v^{n,a}(z)&=& N_{V^{a,n}} v_U^{a,n} (z) \, \, , \, \,
a^{n,a}(z) = N_{A^{a,n}} a_U^{a,n} (z), \label{Normwf}
\eeqa
with the normalization constants defined by the integrals
\beqa
N_{\pi_V^{a,n}}^{-2} &=& \int \frac{dz}{z} \Big \{ ( \partial_z  \tilde \phi_U^{a,n} (z) )^{2} \cr
&+& \beta^{a}_{V}(z) (  \tilde \pi_U^{a,n}(z) -  \tilde \phi_{U}^{a,n}(z) )^{2}  \Big \} \, , \cr
N_{\pi^{a,n}}^{-2} &=& \int \frac{dz}{z} \Big \{ ( \partial_z   \phi_U^{a,n} (z) )^{2} \cr
&+& \beta^{a}_{A}(z) (   \pi_U^{a,n}(z) -   \phi_{U}^{a,n}(z) )^{2}  \Big \} \, , \cr
N_{V^{a,n}}^{-2} &=& \int \frac{dz}{z} (v_U^{a,n}(z) )^{2} \, , \cr
N_{A^{a,n}}^{-2} &=& \int \frac{dz}{z} (a_U^{a,n}(z) )^{2} \, .
\label{NormConst}
\eeqa
The spectrum of vector mesons and scalar mesons is obtained by solving Eqs.~(\ref{ModeEqsVector}) and
imposing the Neumann boundary conditions at the hard wall $z=z_0$. Similarly, the spectrum of axial vector
mesons and pseudoscalar mesons is obtained by solving Eqs.~(\ref{ModeEqsAxial}) and imposing Neumann
boundary conditions at the hard wall.

On the other hand, using Eqs.~(\ref{Asymptotics}) and (\ref{Normwf}) we find that the meson decay
constants, defined in Eqs.~(\ref{gV})-(\ref{fpi}), are determined by the normalization constants through
the relations
\beqa
f_{\pi_{V}^{a,n}} &=&\frac{2}{g_{5}}N_{\pi_V^{a,n}} \, \, , \, \,
f_{\pi^{a,n}} = \frac{2}{g_{5}}N_{\pi^{a,n}} \, , \cr
g_{V^{a,n}} &=& \frac{2}{g_{5}} N_{V^{a,n}} \, \, , \, \,
g_{A^{a,n}} = \frac{2}{g_{5}} N_{A^{a,n}} \, .
\eeqa
Having described the procedure for finding the meson spectrum and decay constants now we describe
how we fit the parameters of our model, namely the quark masses $m_u, m_s, m_c$ , quark condensates
$\sigma_u , \sigma_s, \sigma_c$ and the position of the hard wall $z_0$. We choose to fit the parameter
$z_0$ using only the mass of the $\rho$ meson, since that observable does not depend on any other parameter.
We find $z_0^{-1} = 322.5 {\rm MeV}$. Then we proceed with a global fit for the quark masses and quark
condensates using 10 observables, namely the light meson masses $(m_{\pi},m_{a_1})$, the strange meson
masses $(m_{K},m_{K^\ast},m_{K_1},m_{K_{0}^{*}})$ and the charmed meson masses
$(m_{D},m_{D^\ast},m_{D_s},m_{D^{*}_s})$. Note that we have included the scalar meson $K_{0}^{*}$,
which is associated with flavor symmetry breaking. Numerically, we find the best global fit for the
parameters  $m_{u}=9$ MeV, $m_{s}=190$ MeV and $m_{c}=1560$ MeV for the quark masses;
$\sigma_{u}=(198 \text{MeV} )^{3}$, $\sigma_{s}=(205 \text{MeV})^{3}$ and $\sigma_{c}=(280 \text{MeV})^{3}$
for the quark condensates. In Table \ref{table:fit} we compare the model fit to the observables with
their experimental values.

\begin{table}[!htb]
\caption{Global fit to masses of eleven selected mesons. The mass of the $\rho$ meson
was fit separately using $z_0$.} % title of Table
%\centering % used for centering table
\begin{ruledtabular}
\begin{tabular}{c|rr} % centered columns (4 columns)
Mass & Model (MeV) & Measured (MeV) \\  % inserts table
%heading
\hline % inserts single horizontal line
 % inserting body of the table
$m_{\rho}$ & 775.6 & 775.3 $\pm$ 0.3 \cite{Olive:2016xmw} \\
$m_{\pi}$ & 142.5 & 139.6 \cite{Olive:2016xmw} \\
$m_{a_1}$ & 1232 &  1230 $\pm$ 40  \cite{Olive:2016xmw} \\
$m_{K}$  & 489.2 & 493.7 \cite{Olive:2016xmw} \\
$m_{K^\ast}$  & 803.7 & 891.7 $\pm$ 0.3 \cite{Olive:2016xmw} \\
$m_{K_1}$ & 1359 &   1272 $\pm$ 7 \cite{Olive:2016xmw} \\
$m_{K_{0}^{*}}$ & 674.9 & 682 $\pm$ 29 \cite{Olive:2016xmw} \\
$m_{D}$  & 1831 & 1870 \cite{Olive:2016xmw}\\
$m_{D_{s}}$  & 1987 & 1968 \cite{Olive:2016xmw} \\
$m_{D^{*}}$  & 2161 & 2010 \cite{Olive:2016xmw} \\
$m_{D_{s}^{*}}$  & 2006 & 2112 \cite{Olive:2016xmw} \\
% [1ex] adds vertical space
%\hline %inserts single line
\end{tabular}
\end{ruledtabular}
\label{table:fit} % is used to refer this table in the text
\end{table}

Note that the fit works very well for the isospin and strange sectors, as already known from
previous works. The extension of the model to the charm sector also gives a reasonably good fit
of properties of heavy-light mesons, like the $D$ and $D^\ast$ mesons.

\begin{table}[!htb]
\caption{Set of predictions for masses and decay constants, compared to experimental or lattice data.
The measured value for $f_D$ and $f_{D_s}$, taken from Ref.~\cite{Olive:2016xmw}, are averages from
lattice QCD results. The other measured values are taken from experimental data.} % title of Table
%\centering % used for centering table
\begin{ruledtabular}
\begin{tabular}{c|rr} % centered columns (4 columns)
Observable & Model (MeV) & Measured (MeV) \\ % inserts table
%heading
\hline % inserts single horizontal line
 % inserting body of the table
$f_{\pi}$  & 84.41 & 92.07 $\pm$ 1.2 \cite{Olive:2016xmw} \\
$g_{\rho}^{1/2}$ & 329.3 & 345 $\pm$ 8 \cite{Donoghue:1992dd}  \\
$g_{a_1}^{1/2}$ & 440.9 & 420 $\pm$ 40 \cite{Isgur:1988vm} \\
$f_{K}$  & 98.14 & 110 $\pm$ 0.3 \cite{Olive:2016xmw} \\
$g_{K^\ast}^{1/2}$ & 331 & - \\
$g_{K_1}^{1/2}$ & 478.3 & -  \\
$f_{K_0^*}$ & 33.48 & - \\
$m_{D_1}$ & 2500 & 2423 $\pm$ 2   \cite{Olive:2016xmw}  \\
$m_{D^\ast_0}$ & 1704 & 2318 $\pm$ 29 \cite{Olive:2016xmw}\\
$m_{D^\ast_{0s}}$ & 1547 & 2318 $\pm$ 1 \cite{Olive:2016xmw} \\
$f_{D}$  & 186.7 & 149.8 $\pm$ 0.8 \cite{Olive:2016xmw} \\
$f_{D_{s}}$  & 195 & 176.1 $\pm$ 0.8 \cite{Olive:2016xmw}\\
$g_{D^\ast}^{1/2}$ & 572.7 & - \\
$g_{D^\ast_s}^{1/2}$ & 546.4 & - \\
$g_{D_1}^{1/2}$ & 722 & - \\
$f_{D^\ast_0}$ & 159 &  - \\
$f_{D^\ast_{0s}}$ & 150.5 & - \\
\end{tabular}
\end{ruledtabular}
\label{table:predictions} % is used to refer this table in the text
\end{table}

Once we have fitted the parameters of the model, we are able to make predictions.
In Table (\ref{table:predictions}) we show a set of predictions for masses and decay constants.
In the cases where experimental or lattice data is available the measured values are
presented. Regarding the masses $m_{D^\ast_0}$ and $m_{D^\ast_{0s}}$, the large difference between
the model prediction and the experimental values~\cite{Olive:2016xmw} may be related to the not
clear distinction between the ground and excited states.

\begin{table}[!htb]
\caption{Couplings for the pseudoscalars mesons with the $\rho$ meson and their excitations} % title of Table
%\centering % used for centering table
\begin{ruledtabular}
\begin{tabular}{c|cccccc} % centered columns (4 columns)
$n$ & 0 & 1 & 2 & 3 & 4  \\ [0.5ex] % inserts table
\hline % inserts single horizontal line
$g_{\rho^n \pi \pi}$     & 4.9144 & 1.6019 & -0.8511 & 0.0325 & 0.0231   \\ % inserting body of the table
$g_{\rho^n \rho \rho}$   & 6.8634 &-1.9971 & 0.022 &-0.0025 & 0.0005    \\ % inserting body of the table
$g_{\rho^n K K }$     & 2.2163 & 1.1512 & -0.377 & -0.0241 & 0.0103   \\ % inserting body of the table
$g_{\rho^n K^* K^*}$   & 3.4246  & -0.981  & 0.0013  & 0.0003 & -0.0002     \\ % inserting body of the table
$g_{\rho^n D D}$         & 1.103 & 1.8591 & 0.8386 & 0.0327 & -0.0564    \\ % inserting body of the table
$g_{\rho^n D^{*} D^{*}}$ & 2.1431 & 1.5691 & -0.4344 &-0.3573 &-0.0467   \\ % inserting body of the table
\end{tabular}
\end{ruledtabular}
\label{table:rhocoupling} % is used to refer this table in the text
\end{table}

Now we move to the triple meson couplings defined in the effective Lagrangians (\ref{LVpipi}) and (\ref{LVVV}).
Using the dictionary  (\ref{gVVVv2},~\ref{gpiVpiv2}) and the relations (\ref{RealCouplings}) we find the couplings $g_{\rho^n \pi \pi}$, $g_{\rho^n \rho \rho}$, $g_{\rho^n K K}$, $g_{\rho^n K^* K^*}$, $g_{\rho^n D D}$ and
$g_{\rho^n D^* D^*}$. The results are shown in Table \ref{table:rhocoupling} where we notice an interesting
feature taking place. The triple coupling $g_{\rho^n D D}$, involving the heavy-light pseudoscalar charmed
mesons, does not decrease with $n$ in the same way as the  triple couplings  $g_{\rho^n \pi \pi}$ and
$g_{\rho^n K K}$, involving light and strange pseudoscalar mesons respectively. The same behavior appears
in the triple coupling $g_{\rho^n D^\ast D^\ast}$, involving vectorial charmed mesons,  when compared to
the triple couplings $g_{\rho^n \rho \rho}$ and $g_{\rho^n K^* K^*}$, involving vectorial light and strange
mesons respectively. In the case of $g_{\rho^n D D}$ we see that actually the first resonance $\rho^{n=1}$
couples stronger than the fundamental $\rho^{n=0}$.

\begin{table}[h!]
\caption{ SU(4) flavor symmetry breaking in our model} % title of Table
%\centering % used for centering table
\begin{ruledtabular}
\begin{tabular}{c|cc} % centered columns (4 columns)
Ratios  & $SU(4)$ symmetry & Model   \\  % inserts table %heading
\hline
$\frac{2g_{\rho K K}}{g_{\rho \pi \pi}}$ & 1 & 0.902 \\ % inserting body of the table
$\frac{2g_{\rho K^{*} K^{*}}}{g_{\rho \rho \rho}}$ & 1 & 0.998  \\ % inserting body of the table
$\frac{2g_{\rho D D}}{g_{\rho \pi \pi}}$ & 1 & 0.449 \\ % inserting body of the table
$\frac{2g_{\rho D^{*} D^{*}}}{g_{\rho \rho \rho}}$ & 1 & 0.625  \\ % inserting body of the table
\end{tabular}
\end{ruledtabular}
\label{table:su4breaking}
\end{table}

Due to flavor symmetry breaking, through the different values for the quark masses and condensates, we expect a violation of  $SU(4)$ relations given in~(\ref{su4relation1})-(\ref{su4relation2}). We compare in Table~\ref{table:su4breaking} our
results with the expectations from the $SU(4)$ flavor symmetric case. We note that we find the trend
$g_{\rho D D} < g_{\rho K K} < g_{\rho \pi \pi}/2$ for the pseudoscalar mesons, which opposite to that found with the QCD sum rules~\cite{Bracco:2011pg}
and Dyson-Schwinger calculations in Ref.~\cite{ElBennich:2011py}, but it agrees with calculations based
$^3P_0$ quark-pair creation model in the nonrelativistic quark model of Refs.~\cite{Krein:2012lra}. For the vectorial mesons we find a similar trend $g_{\rho D^* D^*} < g_{\rho K^* K^*} < g_{\rho \rho \rho}/2$. Note, however, that the coupling $g_{\rho K^* K^*}$ is very close to $g_{\rho \rho \rho}/2$. The reason behind this proximity is that the vector meson spectrum, found from the first equation in (\ref{ModeEqsVector}), depends more on the condensate difference than the mass difference appearing in $\beta_V^a(z)$. Since the strange and light condensates $\sigma_s$ and $\sigma_u$ are very close to each other the masses and wave functions of the $\rho$ and $K^*$ are very similar.

\begin{table}[!htb]
%\centering % used for centering table
\caption{Contributions of the five first states to the EM form factors of mesons at $Q^{2}=0$.} % title of Table
\begin{ruledtabular}
\begin{tabular}{c|ccccc} % centered columns (4 columns)
$n$ & 0 & 1 & 2 & 3 & 4 \\  % inserts table
\hline % inserts single horizontal line
$\frac{g_{\rho^n}}{m_{\rho^n}^2}  g_{\rho^n \pi \pi} $
& 0.886 & 0.192 & -0.082 & 0.003 & 0.002
\\ % inserting body of the table
$\frac{g_{\rho^n}}{m_{\rho^n}^2}  g_{\rho^n \rho \rho}$
& 1.237 & -0.239 & 0.002 & 0.000 & 0.000
\\ % inserting body of the table
$2 \frac{g_{\rho^n}}{m_{\rho^n}^2} g_{\rho^n K K}$
& 0.799  & 0.276 & -0.072 & -0.004 & 0.002
\\ % inserting body of the table
$2 \frac{g_{\rho^n}}{m_{\rho^n}^2} g_{\rho^n K^{*} K^{*}}$
& 1.235 & -0.235 & 0.000 & 0.000 & 0.000  \\ %
$2 \frac{g_{\rho^n}}{m_{\rho^n}^2} g_{\rho^n D D}$
& 0.398 & 0.446 & 0.161 & 0.005 & -0.008
\\ % inserting body of the table
$2 \frac{g_{\rho^n}}{m_{\rho^n}^2} g_{\rho^n D^{*} D^{*}}$ & 0.773
& 0.376 & -0.083 & -0.059 & -0.007 \\ % inserting body of the table
\end{tabular}
\end{ruledtabular}
\label{table:PFFDFF0} % is used to refer this table in the text
\end{table}

Finally, we show our results for the meson (elastic) EM form factors defined in the previous section.
Using the couplings obtained in Table \ref{table:rhocoupling} and evaluating the expressions
in Eqs.~(\ref{PionFormFactor})-(\ref{DFormFactor}) we obtain a series expansion for the EM form factors
of the pion, kaon and $D$ meson. In a similar fashion, we take the coupling in Table \ref{table:rhocoupling} and
evaluate the expressions in Eqs.~(\ref{rhoFormFactor})-(\ref{D*FormFactor}) to obtain a series expansion
for the EM form factors of the $\rho$, $K^*$ and $D^*$ mesons. In both cases, a good convergence is achieved
after considering 5 states (4 resonances $\rho^n$ besides the fundamental $\rho^{n=0}$). This is explicitly shown
in Table \ref{table:PFFDFF0} for the case $Q^2=0$.

\begin{figure}[!htb]
\begin{center}
\includegraphics[scale=0.7]{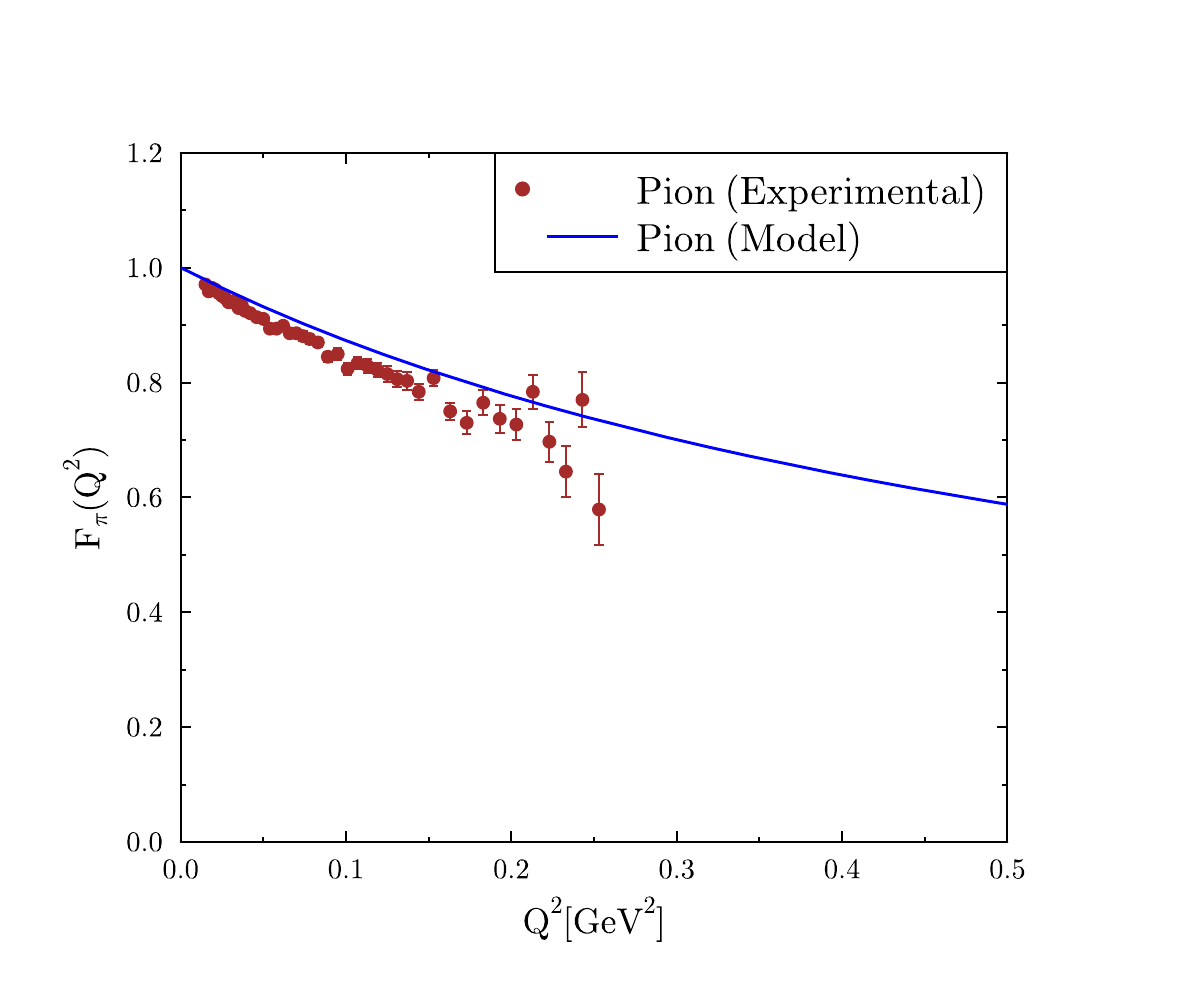}
\end{center}
\vspace{-0.5cm}
\caption{Electromagnetic form factor of the pion (solid line) compared to experimental data \cite{Amendolia:1986wj} (points with error bars).}
\label{PlotPionFormFactor}
\end{figure}

\begin{figure}[!htb]
\begin{center}
\includegraphics[scale=0.7]{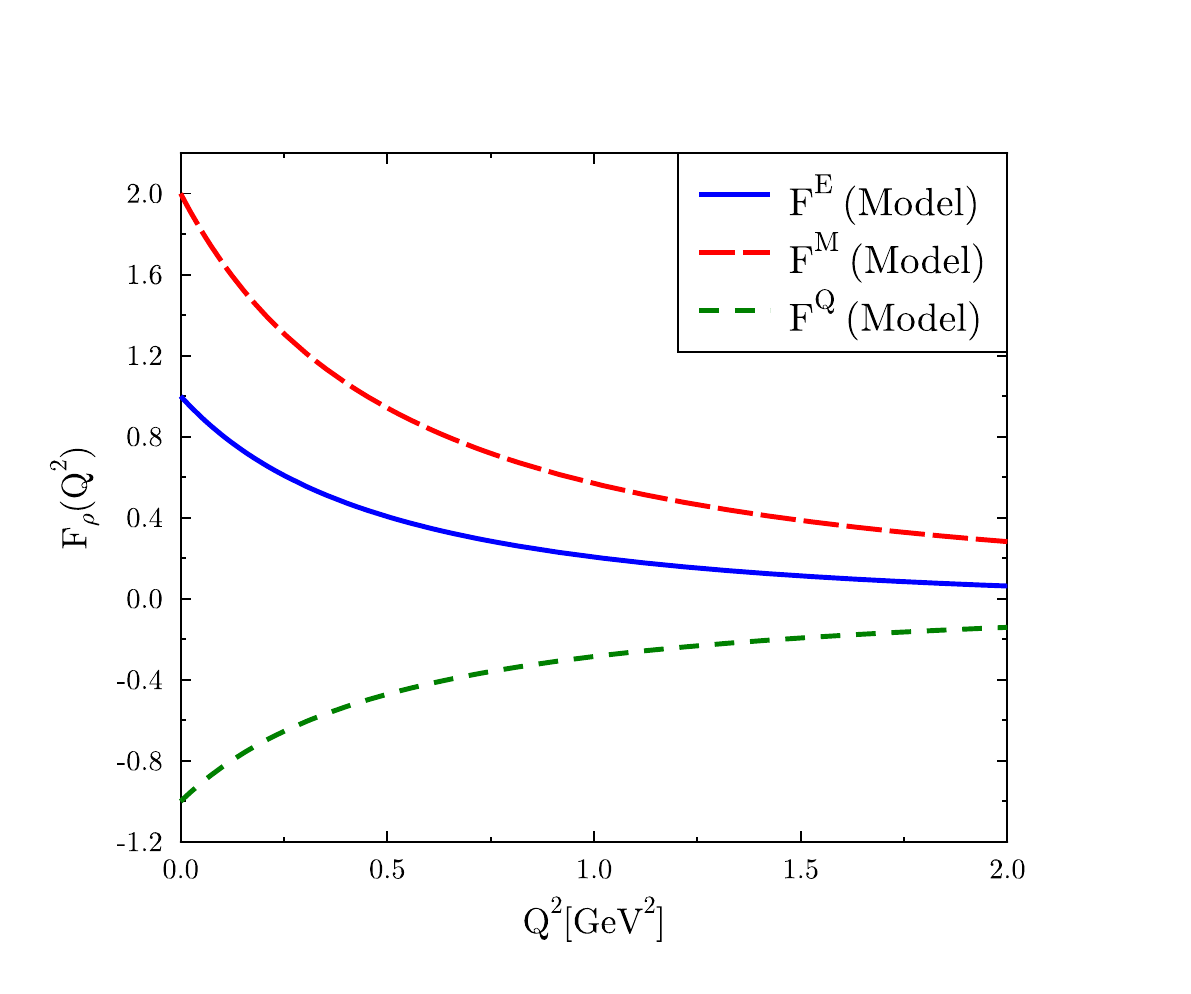}
\end{center}
\vspace{-0.5cm}
\caption{The electric (blue), magnetic (red) and quadrupole (green) EM form factors of the $\rho$ meson.}
\label{PlotrhoFormFactor}
\end{figure}

Table~\ref{table:PFFDFF0} reveals again a clear distinction between the light mesons and charmed mesons.
In the former, vector meson dominance (VMD) is a good approximation, whereas in the case of the latter
the EM form factors receive substantial contribution from the first $\rho$ resonance. This is a nice
example of generalized vector meson dominance (GVMD) in EM form factors. This is consistent with
Ref.~\cite{Bramon:1972vv},  where the authors claimed that the radial excitations of the $\rho$ meson
are important for EM  form factors of nonzero spin hadrons, as nucleons.

\begin{figure}[!htb]
\begin{center}
\includegraphics[scale=0.7]{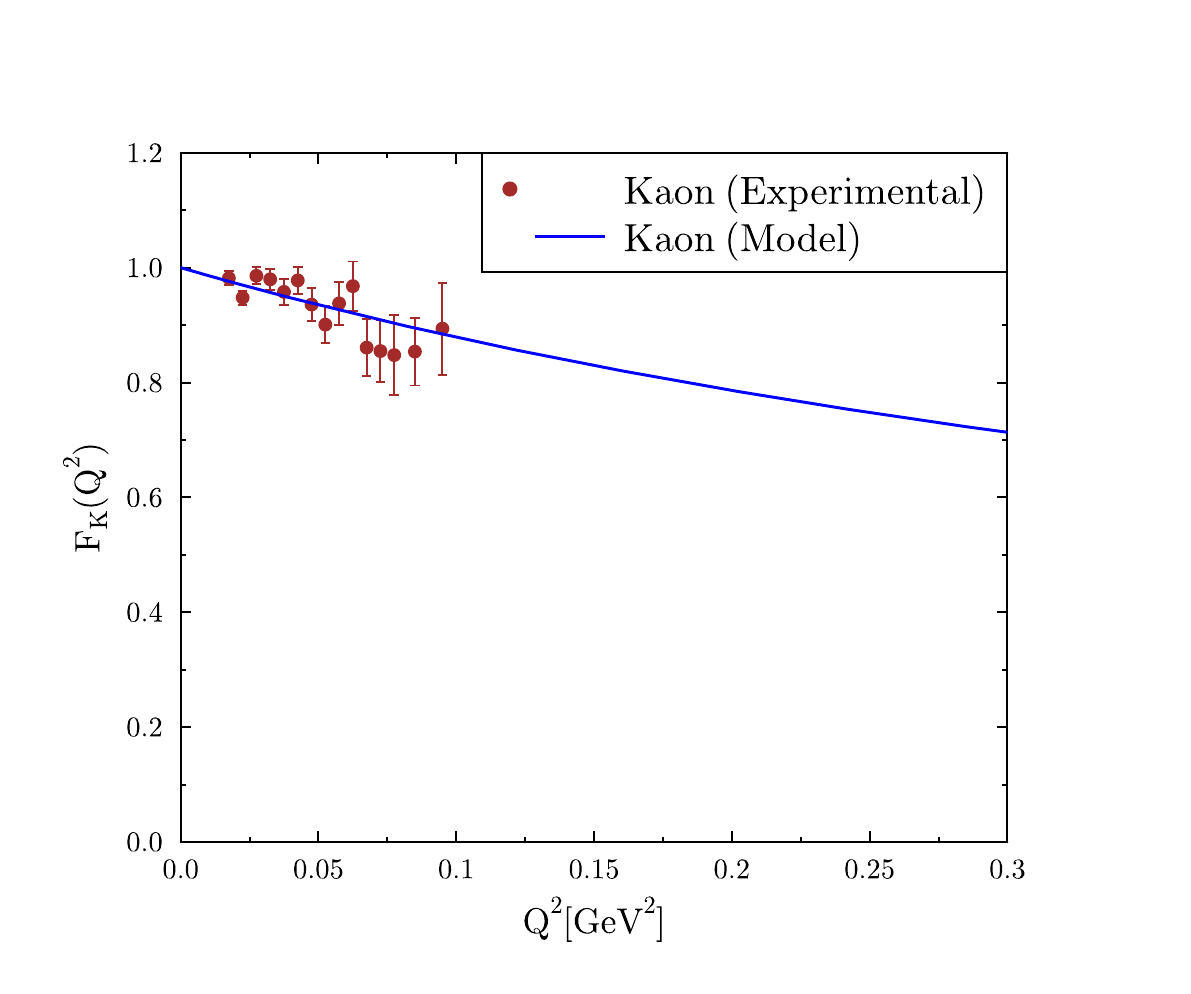}
\end{center}
\vspace{-0.5cm}
\caption{Electromagnetic form factor of the kaon (solid line) compared to experimental data \cite {Amendolia:1986ui} (points with error bars).}
\label{PlotKaonFormFactor}
\end{figure}

\begin{figure}[!htb]
\begin{center}
\includegraphics[scale=0.7]{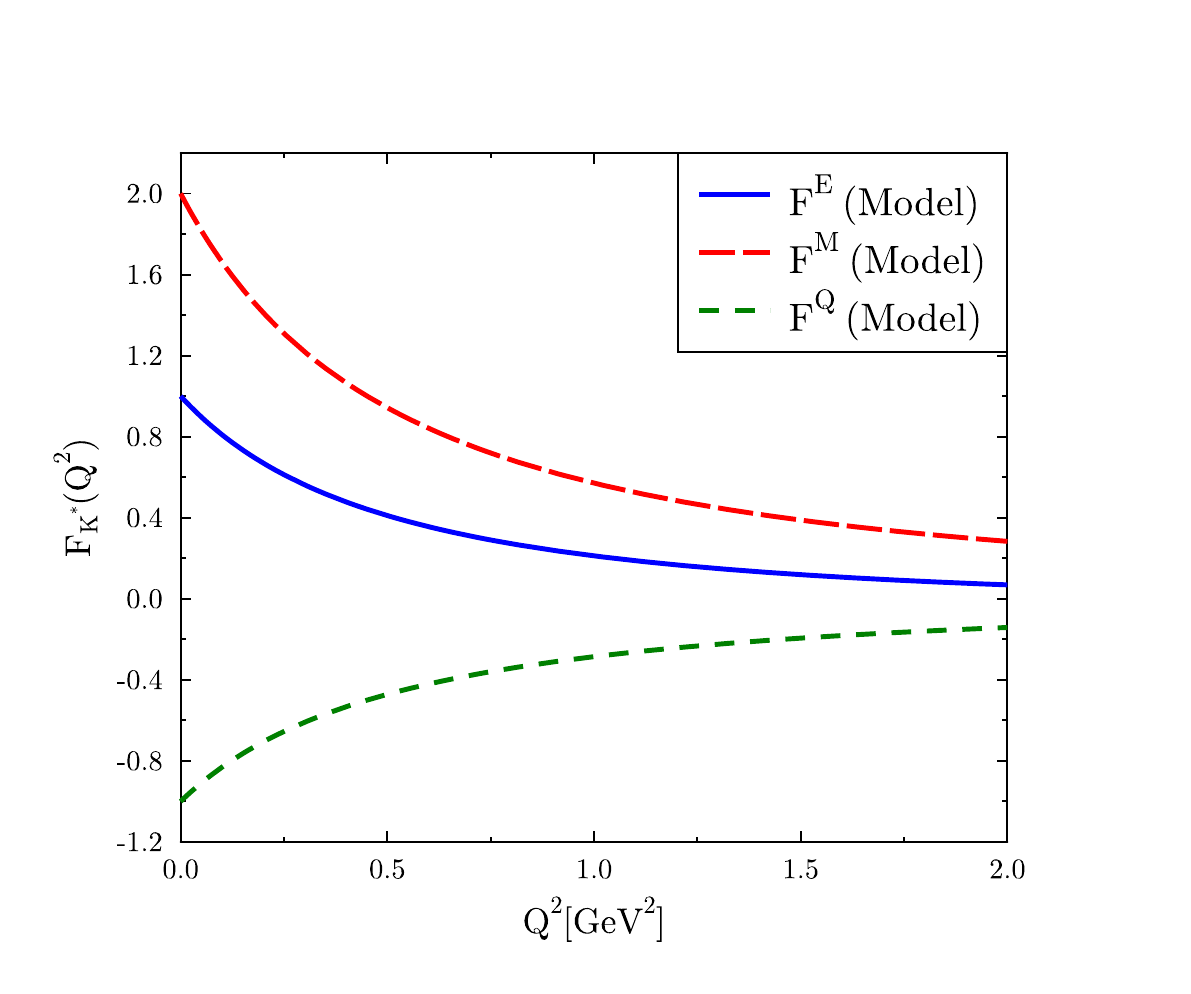}
\end{center}
\vspace{-0.5cm}
\caption{The electric (blue), magnetic (red) and quadrupole (green) EM form factors of the $K^*$ meson.}
\label{PlotKstFormFactor}
\end{figure}

\begin{figure}[!htb]
\begin{center}
\includegraphics[scale=0.7]{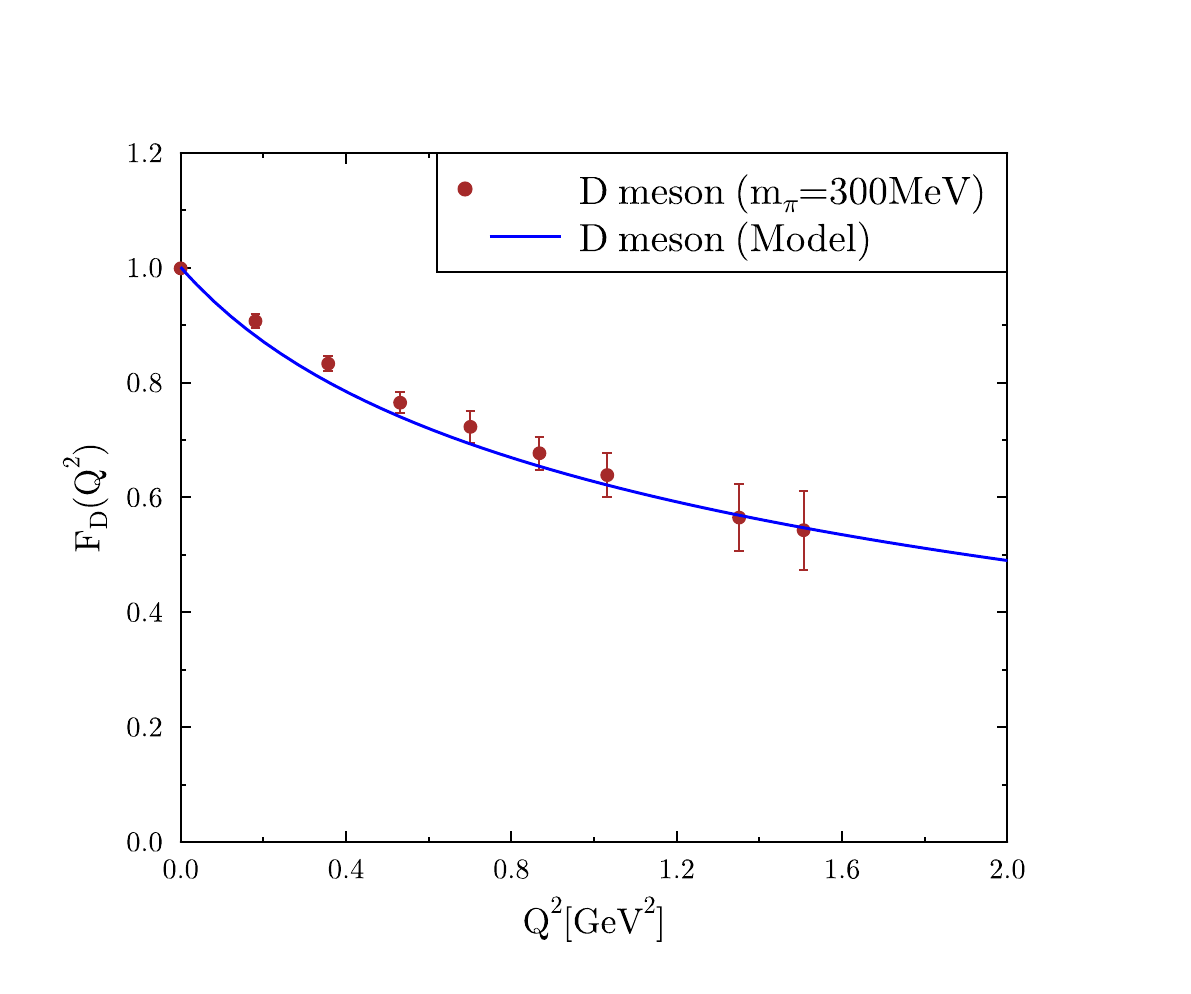}
\end{center}
\vspace{-0.5cm}
\caption{Electromagnetic form factor of the $D$ meson (solid line) compared to lattice QCD data \cite{Can:2012tx} (points with error bars). }
\label{PlotDFormFactor}
\end{figure}

\begin{figure}[!htb]
\begin{center}
\includegraphics[scale=0.7]{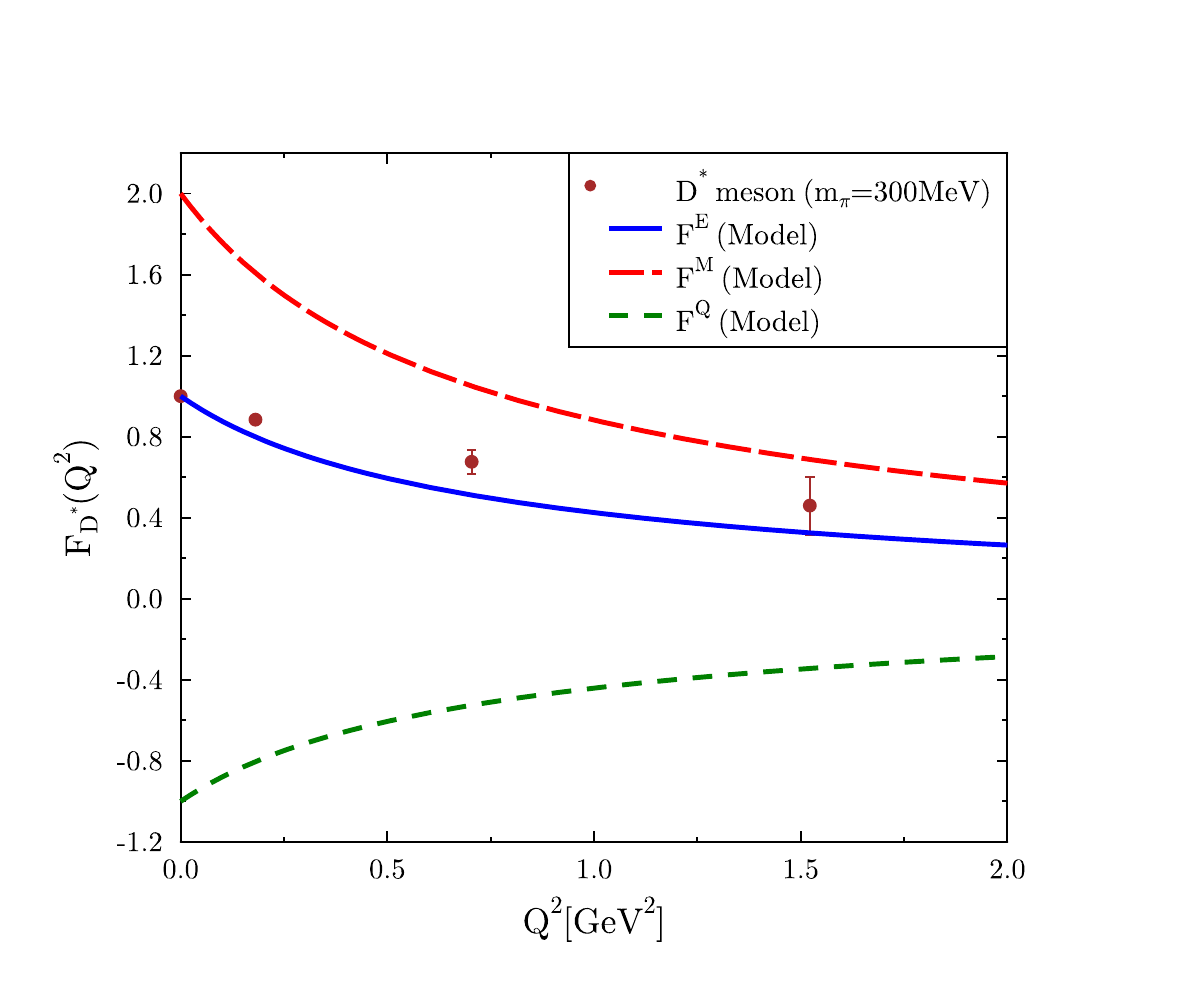}
\end{center}
\vspace{-0.5cm}
\caption{The electric (blue), magnetic (red) and quadrupole (green) EM form factors of the $D^{*}$ meson. The electric form factor $F_{D^*}^E(Q^2)$ (solid blue line) is compared to lattice QCD data \cite{Can:2012tx}
(points with error bars).}
\label{PlotDstFormFactor}
\end{figure}

In Figs.~\ref{PlotPionFormFactor}, \ref{PlotrhoFormFactor}, \ref{PlotKaonFormFactor} and \ref{PlotKstFormFactor} we show our results for the EM form factors of the $\pi$, $\rho$, $K$ and $K^*$ mesons. The pion and kaon EM form factors are compared against experimental data. Previous results for the pion and kaon EM form factors in holographic QCD can be found in \cite{Brodsky:2007hb,Kwee:2007dd} and \cite{Sang:2010kc} respectively. Previous results for the $\rho$ meson EM form factor can be found in \cite{Grigoryan:2007vg}. Finally we show in Figs. \ref{PlotDFormFactor} and \ref{PlotDstFormFactor} our results for the $D$ meson and $D^\ast$ meson EM form factors compared with data from lattice QCD. As promised, we find a reasonable agreement between our model and the lattice results.

\begin{table}[!htb]
\caption{The charge radii of $\pi$, $\rho$, $K$, $K^*$, $D$ and $D^{*}$ meson compared to experiment and lattice QCD.}
% title of Table
%\centering % used for centering table
\begin{ruledtabular}
\begin{tabular}{c|ccc} % centered columns (4 columns)
$\langle r^2 \rangle$ (fm$^2$) & Model & Experiment  & Lattice QCD \\  % inserts table
\hline % inserts single horizontal line
$\pi^{+}$ & 0.35 & 0.45 $\pm$ 0.01 \cite{Olive:2016xmw} & -      \\ % inserting body of the table
$\rho^{+}$ &0.53   & 0.56 $\pm$ 0.04 \cite{Krutov:2016uhy} & -    \\
$K^{+}$ & 0.33 & 0.31 $\pm$ 0.03 \cite{Olive:2016xmw} & -  \\
$K^{*+}$ & 0.52 & - & -  \\
$D^{+}$ & 0.19  & - & 0.14 $\pm$  0.01 \cite{Can:2012tx}  \\
$D^{*+}$ & 0.33  & - & 0.19 $\pm$ 0.02  \cite{Can:2012tx}  \\
% inserting body of the table
\end{tabular}
\end{ruledtabular}
\label{table:chargeradius} % is used to refer this table in the text
\end{table}

At low $Q^2$, we use the relations in Eqs.~(\ref{PSFormFactorLowQ2}) and (\ref{VMchargeradius}) to extract
the charge radii of pseudoscalar and vector mesons. In Table \ref{table:chargeradius} we compare our results
for the light mesons against experimental data and those for the charmed mesons against lattice QCD data.

\begin{figure}[!htb]
\begin{center}
\includegraphics[scale=0.7]{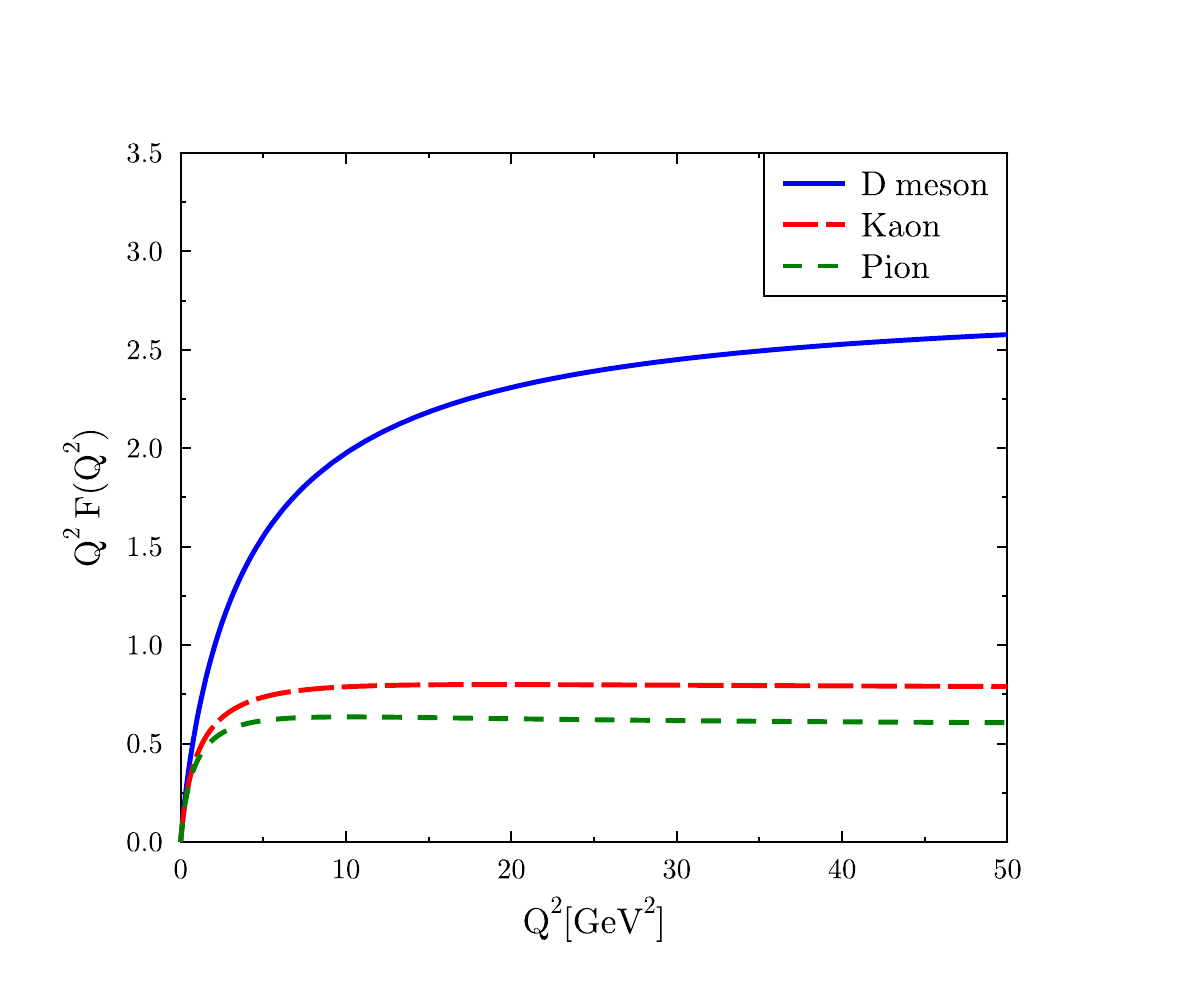}
\end{center}
\vspace{-0.5cm}
\caption{Large $Q^2$ behavior for the $D$-meson, kaon and pion electromagnetic form factor}
\label{PlotPSFormFactorHighQ2}
\end{figure}

\begin{figure}[!htb]
\begin{center}
\includegraphics[scale=0.7]{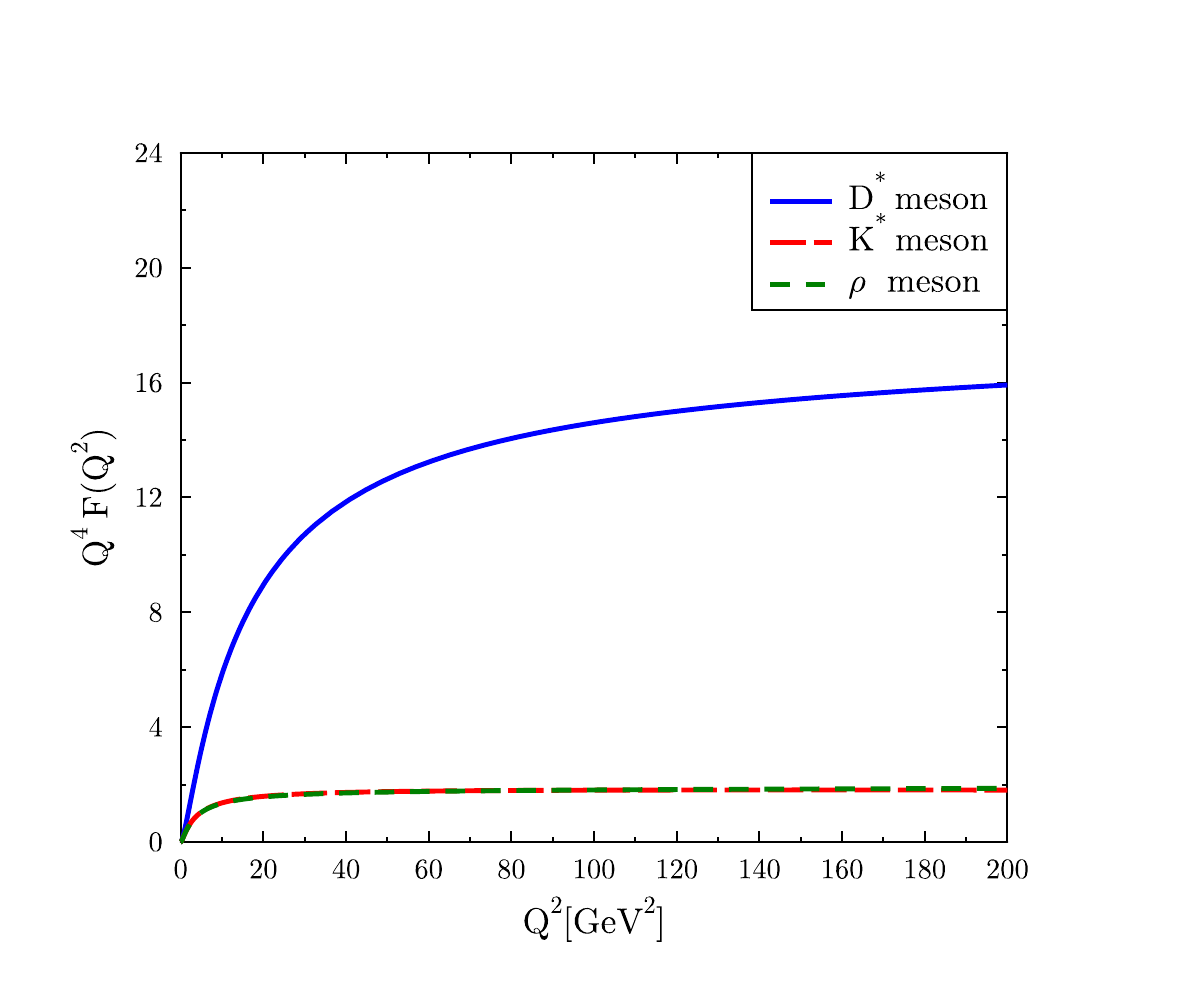}
\end{center}
\vspace{-0.5cm}
\caption{Large $Q^2$ behavior for the $D^{*}$-meson, $K^*$-meson and $\rho$-meson electromagnetic form factor}
\label{PlotVMFormFactorHighQ2}
\end{figure}

We use the expansions in Eqs.~(\ref{PSFormFactorHighQ2}) and (\ref{VMFormFactorHighQ2}) to obtain the $Q^2$
dependence of the EM form factors; the results are shown in Figs.~\ref{PlotPSFormFactorHighQ2} and
\ref{PlotVMFormFactorHighQ2}. For the case of  pseudoscalar mesons, we find the behavior $F_{\pi^a}(Q^2) \sim Q^{-2}$. For the vector meson EM form factors we find that $F_{V^a}(Q^2) \sim Q^{-4}$. Both results are
consistent with perturbative QCD expectations and conformal symmetry in the UV.

%%%%%%%%%%%%%%%%%%%%%%%%%%%%%%%%%%%%%%%%%%%%%%%%%%%%%%%%%%%%%%%%%%%%%%%%%%%%%%%%%%%%%%%%%%%
\section{Conclusions}
\label{Sec:concl}

We have extended the two-flavor hard-wall holographic model of Ref.~\cite{Erlich:2005qh} to four
flavors. By fitting the seven parameters of the model, which are three quark masses, three condensates
and the hard-wall scale $z_0$, to eleven selected meson masses, the model provides a good description
of weak decay constants of more than a dozen of light and strange and charmed mesons. We have also
investigated the effects of flavor symmetry breaking on three-meson couplings and form factors.
In particular, we have made predictions for the strong couplings $g_{\rho^{n} \pi \pi}$, $g_{\rho^{n} \rho \rho}$,  $g_{\rho^{n} K K}$, $g_{\rho^{n} K^{*} K^{*}}$,  $g_{\rho^{n} D D}$ and $g_{\rho^{n} D^{*}D^{*}}$. Moreover, using our results for those couplings we have been able to
evaluate the $\pi$, $\rho$, $K$, $K^*$, $D$ and $D^{*}$ electromagnetic form factors. For the $D$ and $D^{*}$  electromagnetic form factors we found a reasonable agreement with the lattice QCD results of Ref.~\cite{Can:2012tx}.

Our results for the  couplings involving the ground-state $\rho$ meson and the charmed mesons,
namely $g_{\rho DD}$ and $g_{\rho D^* D^*}$ are smaller than the $SU(4)$ symmetry values, as shown
in Table~\ref{table:su4breaking}. Our result $g_{\rho DD}=1.103$ is also smaller than
predictions based on the VMD model~\cite{Mat98,Lin00a} where $g_{\rho D D} = 2.52 - 2.8$. Moreover,
we found that $g_{\rho DD} < g_{\rho\pi\pi}/2$, which is of the opposite trend to the predictions
based on QCD sum rules~\cite{Bracco:2011pg} and Dyson-Schwinger equations of QCD~\cite{ElBennich:2011py},
but it agrees with that obtained with the  $^3{\rm P}_0$ pair-creation model in the nonrelativistic quark model
of Refs.\cite{{Krein:2012lra},{Fontoura:2017ujf}}. A possible explanation for the discrepancy for the small values
of the couplings is that the electromagnetic form factor of the $D$ meson is a dramatic example where the
VMD approximation is broken and the contribution from the resonances $\rho^n$ can not be neglected. It is
interesting to notice the relation between the breaking of the VMD approximation and the breaking of the
$SU(4)$ symmetry. In a VMD approximation we would find that
\beqa
2 g_{\rho D D} = \frac{m_{\rho}^2}{g_\rho} = g_{\rho \pi \pi} \,.
\label{Universality}
\eeqa
The first equality comes from applying VMD to the $D$ isospin form factor; in our framework, this relation
also comes from the EM form factor. The second equality is the well known VMD result for the pion EM
form factor.  The relation in Eq.~(\ref{Universality}) can be extended to other couplings and it means
that a VMD approximation necessarily implies a universality between the couplings. Interestingly, the
result in Eq.~(\ref{Universality}) for the coupling $g_{\rho DD}$ matches with the $SU(4)$ symmetry
expectations. Then it is reasonable to interpret a dramatic breaking of the $SU(4)$ flavor symmetry
in terms of a dramatic breaking of the VMD approximation, which is exactly what we have found for the
charmed mesons $D$ and $D^{*}$.

We finish this paper by reiterating our earlier remarks on the applicability of our model. Our holographic
QCD model is based on an extensions of a light-flavor chiral Lagrangian, which should be adequate
to describe heavy-light mesons, as the internal structure of these mesons is governed by essentially the
same nonperturbative physics governing the internal structure of light mesons, which occurs at the scale
$\Lambda_{\rm QCD}$.   On the other hand, the internal structure of heavy-heavy mesons, as the
$\psi$ and $\eta_c$ mesons, is governed by short-distance physics at the scale of the heavy quark
mass. An appropriate holographic description of such mesons most likely requires the inclusion of
long open strings. In that scenario, it should be possible, in particular, to describe the non-relativistic
limit of heavy quarks where a spin-flavor symmetry emerges~\cite{Manohar:2000dt}. Although there have
been some interesting top-down~\cite{Erdmenger:2006bg,Erdmenger:2007vj,Hashimoto:2014jua,Liu:2016iqo} and
bottom-up~\cite{Braga:2015lck,Liu:2016urz} proposals, a realistic model for heavy-heavy mesons
remains a challenge in holographic QCD.

In holographic QCD, it is assumed that the quark mass coefficient $m_q$ in the near boundary expansion of the classical field $X_0(z)$ behaves as the source of the operator $ \bar{q}(x)q(x)$. Then the holographic dictionary leads to conclude that the parameter $\sigma$ is also in one to one correspondence with the vacuum expectation value (v.e.v.) $\langle \bar{q}(x)q(x) \rangle$. This matching, however, is ambiguous, as discussed in Ref. \cite{Cherman:2008eh}, because $\langle \bar{q}(x)q(x) \rangle$ is actually a scale-dependent quantity whereas $m_q$ and $\sigma$ are obtained from a global fit to the meson spectrum. This issue actually becomes exacerbated as the quark mass increases.

In QCD, the quantity $\langle \bar{q}(x)q(x) \rangle$ is identified with the trace of the quark propagator $S$, i.e. $\langle \bar{q}(x)q(x) \rangle = - {\rm Tr} \, S(x-x) 
= - {\rm Tr} \, S(0)$. It contains a nonperturbative, low-energy contribution from a dynamical component of chiral symmetry breaking, and an essentially perturbative contribution due to the explicit chiral symmetry breaking driven by the quark mass. In the heavy quark mass limit, the perturbative contribution dominates and the nonperturbative contribution goes to zero. For that reason, and to make contact 
with the traditional definition of the quark condensate in  QCD sum rules \cite{Reinders:1984sr}, in lattice simulations the perturbative contribution is subtracted; see e.g. \cite{McNeile:2012xh}. Interestingly, 
the authors of Ref. \cite{McNeile:2012xh} found, after subtracting the 
perturbative contribution, that the strange quark condensate at the $\overline{MS}$ scale of 2~{\rm GeV}  is larger than that of the light quarks. So far there are no such lattice calculations for the charm and bottom quark, 
but calculations within the framework of Dyson-Schwinger equations \cite{Chang:2006bm} find that the nonperturbative component of chiral symmetry breaking decreases 
with increasing current-quark mass, as expected. 

Our results, obtained from a global fit to the meson spectrum, indicate that $\sigma$ 
increases with $m_q$. Although, as discussed above, the relation between $\sigma$ and the 
QCD v.e.v. $\langle \bar{q}(x)q(x) \rangle$ is far from clear, one could assume that relation
as being strictly one-to-one and conclude that $\langle \bar{q}(x)q(x) \rangle$ increases with the quark mass unless a perturbative subtraction is also implemented in the holographic model. For the case of the charm quark this means that a large value for $\sigma_c$ does not necessarily imply a large charm quark condensate. There is an additional issue that requires further study. In QCD, the v.e.v. of the operator $\bar q q$, with canonical dimension 
$\Delta=3$, is expected to acquire a large anomalous dimension in the infrared. In our holographic 
model, we have made the ad hoc approximation of keeping the same canonical dimension for $\langle \bar q q \rangle$. If we take into account anomalous dimension effects, 
corrections to $m_q$ and $\sigma$ are expected. We hope to pursue this line of research in the near future.

It is also important to bear in mind that results from a holographic QCD approach are supposedly
referring to leading-order in an expansion of $1/N_c$ and in the large 't Hooft coupling $\lambda=g_{YM}^2 N_c$.
As such, loop corrections for the hadronic propagators and vertices are not taken into account.
The $1/N_c$ and/or $1/\lambda$ corrections to the effective chiral-flavor Lagrangians
in holographic QCD is a fascinating open problem and deserves further studies.

%%%%%%%%%%%%%%%%%%%%%%%%%%%%%%%%%%%%%%%%%%%%%%%%%%%%%%%%%%%%%%%%%%%%%%%%%%%%%%%%%%%%%%%%%%%
\section*{Acknowledgements}

Work partially supported by Funda\c{c}\~ao de Amparo \`a Pesquisa do Estado de
S\~ao Paulo (FAPESP), Grants No. 2015/17609-3 (A.B.-B.) and 2013/01907-0 (G.K.),  
Conselho Nacional de Desenvolvimento Cient\'{\i}fico e Tecnol\'ogico (CNPq), Grant
No. 305894/2009-9 (G.K.) and Coordena\c{c}\~ao de Aperfei\c{c}\~oamento de Pessoal de N\'{\i}vel
Superior (CAPES) for a doctoral fellowship (C.M).  A.B-B also 
acknowledges partial financial support from the  grant 
CERN/FIS-NUC/0045/2015. The authors also thank K.U.~Can for providing them
the lattice data for the electromagnetic form factors of Ref.~\cite{Can:2012tx}.

%%%%%%%%%%%%%%%%%%%%%%%%%%%%%%%%%%%%%%%%%%%%%%%%%%%%%%%%%%%%%%%%%%%%%%%%%%%%%%%%%%%%%%%%%%%

\end{document}